\documentclass[aps,pra,reprint,superscriptaddress]{revtex4-2}

\usepackage{microtype}
\usepackage{graphicx}  
\usepackage{amsmath,amssymb}  
\usepackage{bm}         
\usepackage{dcolumn}   
\usepackage{xcolor}
\usepackage{mathtools}
\usepackage[hidelinks]{hyperref}   
\usepackage[capitalise]{cleveref}

\renewcommand{\Re}{\operatorname{Re}}
\renewcommand{\Im}{\operatorname{Im}} 
\newcommand{\Tr}[2][]{\operatorname{Tr}_{#1}\left(#2\right)}
\newcommand{\mean}[1]{\mathbb{E}\left[#1\right]}

\DeclareMathOperator{\sgn}{sgn}
\DeclarePairedDelimiter\norm{\lVert}{\rVert}

\begin{document}

\title{Fibonacci Steady-States and Persistent Oscillations in an Ordered Multimode Dicke Model}

\author{Miriam J. Leonhardt}
\affiliation{Clarendon Laboratory, University of Oxford, Parks Road, Oxford OX1 3PU, United Kingdom}

\author{Kai Müller}
\affiliation{Institute of Theoretical Physics, TUD Dresden University of Technology, 01062 Dresden, Germany}

\author{Oliver Lunt}
\affiliation{Clarendon Laboratory, University of Oxford, Parks Road, Oxford OX1 3PU, United Kingdom}

\author{Andrew J. Daley}
\affiliation{Clarendon Laboratory, University of Oxford, Parks Road, Oxford OX1 3PU, United Kingdom}

\date{\today}

\begin{abstract}
Ultracold atoms in multimode optical cavities provide a rich testbed for many-body phenomena enabled by light-mediated interactions. Recent experiments include realizations of spin glasses and associative memories, as described by multimode Dicke models with disordered couplings. However, the properties of multimode Dicke models with ordered coupling geometries remain largely unexplored. In this work, we investigate the stable steady-states of the multimode Dicke model with an ordered nearest-neighbor coupling geometry, where $n_c$ atomic clusters are coupled via $n_c-1$ cavity modes. We show that the number of mean-field stable steady-states in the superradiant phase exhibits Fibonacci scaling with the number of atomic clusters, and that a subset of these steady-states exhibit persistent oscillations. Using both the truncated Wigner approximation and the numerically-exact hierarchy of pure states, we further demonstrate that these features of the stable steady-state solutions persist for finite cluster sizes. Ordered multimode Dicke models, such as the nearest-neighbor coupling geometry considered here, are accessible with current experimental technologies and point toward a broader class of strongly interacting dissipative systems with similarly rich behavior.

\end{abstract}

\maketitle
\section{Introduction}
The Dicke model is a paradigmatic model for the collective quantum phenomena enabled by the interaction of light and matter~\cite{hepp_equilibrium_1973,hepp_superradiant_1973,wang_phase_1973}. In particular, it describes the collective interaction between an ensemble of two-level emitters and a single mode of the electromagnetic field. As a function of the light-matter coupling, it exhibits a continuous quantum phase transition between a normal phase and a superradiant phase, respectively characterized by subextensive and extensive occupation of the cavity mode~\cite{kirton_introduction_2019,dimer_proposed_2007,emary_chaos_2003,Baumann2010Apr,Klinder2015Mar,Kroeze2018Oct,Buhler2026Apr}. Recently there has been significant attention devoted to variants of the standard Dicke model, including unbalanced Dicke models~\cite{bhaseen_dynamics_2012,stitely_nonlinear_2020,muller_genuine_2025,kirton_superradiant_2018,gen_Dicke_Singapur}, non-reciprocal Dicke models~\cite{jachinowski_spin-only_2026}, Dicke models with multiple clusters~\cite{bhattacherjee_non-equilibrium_2014,mivehvar_conventional_2024,iemini_dynamics_2024,zhao_collective_2017} and---most pertinent to us---multimode Dicke models~\cite{marsh_entanglement_2024,gopalakrishnan_frustration_2011,buchhold_dicke-model_2013,hosseinabadi_quantum--classical_2024,marsh_multimode_2025,tolkunov_quantum_2007,chiacchio_emergence_2018,moodie_generalized_2018,guo_sign-changing_2019,guo_emergent_2019,marsh_enhancing_2021,link_non-markovian_2022,hosseinabadi_far_2024,fiorelli_signatures_2020,reviewCQED_Francesco}. Many of these variations show interesting steady-state behavior, such as bistability and limit cycles~\cite{stitely_nonlinear_2020,adiv_nonlinear_2026,mivehvar_conventional_2024,iemini_dynamics_2024,bhattacherjee_non-equilibrium_2014}.

The multimode Dicke model, in which multiple ensembles of two-level emitters couple to multiple distinct field modes, was first introduced in~\cite{hepp_superradiant_1973} as a generalization of the Dicke model worthy of further exploration. It has gained traction particularly in the last decade through its experimental implementation in the (nearly) confocal multimode cavities of~\cite{kroeze_directly_2025,marsh_high-capacity_2025,kroeze_high_2023,guo_emergent_2019,vaidya_tunable-range_2018,guo_optical_2021,kollar_adjustable-length_2015,kollar_supermode-density-wave-polariton_2017} and its relevance to spin glass physics~\cite{strack_dicke_2011,gopalakrishnan_frustration_2011,buchhold_dicke-model_2013,kroeze_directly_2025,marsh_entanglement_2024,hosseinabadi_quantum--classical_2024,marsh_multimode_2025,gopalakrishnan_emergent_2009} and associative memories~\cite{marsh_enhancing_2021,marsh_quantum-optical_2024,marsh_high-capacity_2025,fiorelli_signatures_2020,gopalakrishnan_exploring_2012}. For both spin glass physics and associative memories, the effective, cavity mode mediated interactions between different atomic ensembles must have disorder and be sign-changing to allow for the characteristic many metastable local energy minima to occur. Under steepest-descent dynamics, these metastable energy minima form a subset of the stable steady-state solutions of the system~\cite{marsh_enhancing_2021,erba_self-induced_2021}.

\begin{figure*}[!ht]
  \includegraphics[width=0.9\textwidth]{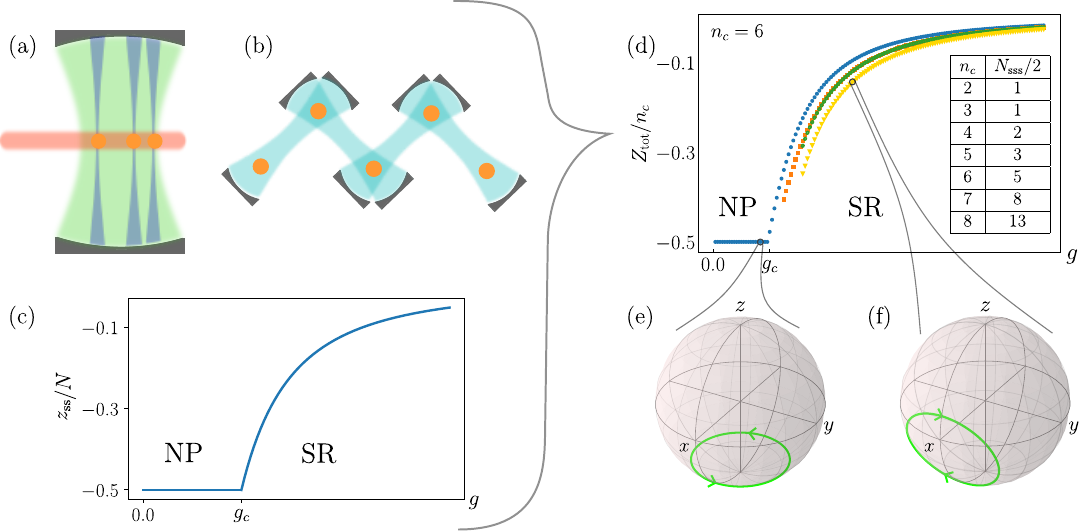}
  \caption{(a) Schematic diagram of a confocal multimode cavity QED apparatus like those in~\cite{marsh_entanglement_2024,kroeze_directly_2025,marsh_high-capacity_2025,kroeze_high_2023,guo_emergent_2019,vaidya_tunable-range_2018,guo_optical_2021}. Multiple atomic spin clusters (orange) are located at distinct positions ($r_i$) in the central plane of the confocal cavity, and are illuminated by the transverse pumping laser (red) causing light to be scattered into the cavity. The cavity field has both a local component (blue), and a non-local component (green) through which the interactions between the spin clusters are mediated. (b) Schematic diagram of intersecting single-mode cavities (cavity fields shown in light blue), with two atomic clusters (orange) in each cavity, arranged to realize the nearest-neighbor coupling geometry of the multimode Dicke model. (c) Stable steady-state solution of the standard, single-mode Dicke model, shown through the $z$-spin component of the atomic cluster, scaled by the number of atoms in the cluster, plotted as a function of the coupling strength $g$. A phase transition from the normal phase (NP) to the superradiant phase (SR) occurs at a critical coupling strength of $g_c$. (d) Stable steady-state solutions of the nearest-neighbor multimode Dicke model shown through the total $z$-spin component of the atomic clusters, scaled by the number of clusters, here $n_c=6$. In the superradiant phase there are multiple distinct pairs of stable steady-state solutions, whose number shows a Fibonacci scaling with the number of clusters $n_c$ (see table). The plot shows $n_{c}=6$, with five solution pairs, two of which take on the same values of $Z_\mathrm{tot}/n_c$. (e), (f) Persistent oscillations in the normal and superradiant phase respectively, shown through a closed orbit on the generalized Bloch sphere of the first cluster.
  }
  \label{fig:Fig1}
\end{figure*}

Motivated by the flexibility now available in experimental cavity QED, we now ask what happens in the case of \emph{ordered} couplings. Compared to their previously studied few-mode counterparts~\cite{moodie_generalized_2018,lang_collective_2017,fan_hidden_2014,wickenbrock_collective_2013}, we find that systems of many cavity modes and many atomic clusters with ordered couplings can exhibit unexpectedly rich steady-state structure. These ordered multimode Dicke models could be experimentally realized by making use of a confocal multimode cavity setup like those in~\cite{kroeze_directly_2025,marsh_high-capacity_2025,kroeze_high_2023,guo_emergent_2019,vaidya_tunable-range_2018,guo_optical_2021,kollar_adjustable-length_2015,kollar_supermode-density-wave-polariton_2017} and carefully choosing the positions of the atomic clusters as depicted in Fig.~\ref{fig:Fig1}(a), or by building cavities that intersect~\cite{leonard_supersolid_2017,baumgartner_stability_2025,leonard_monitoring_2017,morales_coupling_2018}, in order to directly provide a specific geometry, as shown in Fig.~\ref{fig:Fig1}(b). We focus on the initial case of nearest-neighbor coupling in the multimode Dicke model, and characterize its steady-states in the near mean-field regime. This structure is a starting point for other, more complex ordered coupling geometries.

Like the standard single-mode Dicke model, whose stable steady-states are shown in Fig.~\ref{fig:Fig1}(c), the nearest-neighbor multimode Dicke model undergoes a phase transition from the normal phase to the superradiant phase at a critical coupling strength. The mean-field stable steady-state solution has all atomic ensembles de-excited in the normal phase, while in the superradiant phase the atomic ensembles contain excitations (see Fig~\ref{fig:Fig1}(d)).

However, unlike the standard single-mode Dicke model, we find multiple branches of stable steady-state solutions in the superradiant phase of the nearest-neighbor multimode Dicke model. Surprisingly, the number of stable steady-state solutions in the superradiant phase grows with the Fibonacci series, that is, \emph{exponentially}, when the number of clusters $n_{c}$ is increased. 
Furthermore, we show that the mean-field stable steady-state solutions in the normal phase and certain branches of stable steady-state solutions in the superradiant phase correspond to centers of infinitely many periodic orbits on the generalized Bloch spheres of the atomic ensembles, examples of which are depicted in Figs.~\ref{fig:Fig1}(e) and (f) respectively. Such periodic orbits correspond to oscillations in the expectation values of the spin ensembles which, in the mean-field description, persist indefinitely. These unexpected features of multistability and persistent oscillations in the steady-state solutions demonstrate that there is also rich physics in multimode Dicke models with ordered coupling geometries. Using the truncated Wigner approximation~\cite{polkovnikov_phase_2010}, we show that these features of the steady-state solutions can be observed away from the mean-field limit, for experimentally relevant finite-size clusters, and we validate this using the numerically-exact hierarchy of pure states~\cite{suess_hierarchy_2014, hartmann_exact_2017, muller_quantum_2025}.

The remainder of the paper is organized as follows. In Sec.~\ref{sec:model} we introduce the multimode Dicke model and the nearest-neighbor geometry, including the atom-only description used to derive the mean-field equations. Then, in Sec.~\ref{sec:steady-state-slns} we present the mean-field equations and numerically solve for their stable steady-state solutions. We prove the Fibonacci scaling of the number of stable steady-state solutions deep in the superradiant phase and explain why the persistent oscillations observed in both the normal and superradiant phases are stable. We use the truncated Wigner approximation and the hierarchy of pure states method to study the model in Sec~\ref{sec:finite-size-clusters} and confirm that the additional stable steady-states and oscillatory solutions can also be observed for finite cluster sizes. In Sec~\ref{sec:expt_implementations}, we discuss to what extent this model could be implemented experimentally in two different platforms. Finally, in Sec.~\ref{sec:Conc} we summarize the main results and present future perspectives on this work.

\section{Multimode Dicke Model}\label{sec:model}
A multimode generalization to the Dicke model is described by the Hamiltonian (in units with $\hbar=1)$,
\begin{equation}\label{eq:MMC_hamiltonian}
    \hat{H}=\sum_{i=1}^{n_c}\omega_{A_i}\hat{S}_i^z+\sum_{j=1}^{n_m}\omega_{C_j}\hat{a}_j^\dagger\hat{a}_j+\sum_{i=1}^{n_c}\sum_{j=1}^{n_m}\frac{2g_{ij}}{\sqrt{N_i}}(\hat{a}_j+\hat{a}_j^\dagger)\hat{S}_i^x,
\end{equation}
where we consider $n_c$ different clusters of $N_i$ two-level atoms with atomic frequency $\omega_{A_i}$, described by the collective spin operators $\hat{S}^\beta=\sum_{k=1}^N\hat{\sigma}_k^\beta/2$. The clusters couple to $n_m$ different cavity modes, described by the annihilation (creation) operators $\hat{a}_j\ (\hat{a}_j^\dagger)$ and the cavity mode frequencies $\omega_{C_j}$, via varying coupling strengths $g_{ij}$. These strengths depend on the detailed setup of multimode cavity QED experiments ~\cite{marsh_entanglement_2024,marsh_multimode_2025,marsh_enhancing_2021,guo_sign-changing_2019,guo_emergent_2019}, as discussed further in Sec.~\ref{sec:expt_implementations}. The main source of dissipation is photon loss through the cavity mirrors, at a rate $\kappa$, which is modeled using a Lindblad form master equation~\cite{lindblad_generators_1976,gorini_completely_1976},
\begin{equation}\label{eq:MMC_me}
    \dot{\hat{\rho}}=-i[\hat{H},\hat{\rho}]+\kappa\sum_{j=1}^{n_m}D[\hat{a}_j]\hat{\rho},
\end{equation}
where we introduce $D[\hat{\eta}]=2\hat{\eta}\hat{\rho}\hat{\eta}^\dagger-\hat{\rho}\hat{\eta}^\dagger\hat{\eta}-\hat{\eta}^\dagger\hat{\eta}\hat{\rho}$. Other sources of dissipation such as spontaneous emission from the atoms can be made weak relative to $\kappa$ by using a Raman coupling scheme~\cite{marsh_enhancing_2021,dimer_proposed_2007}, and can thus be neglected.

Assuming perfectly degenerate cavity modes ($\omega_{C_j}=\omega_C$) and identical atomic clusters ($\omega_{A_i}=\omega_A$ and $N_i=N$), it was shown in~\cite{marsh_entanglement_2024} that this multimode Dicke model has a nonequilibrium phase transition from the normal phase to a superradiant phase at 
\begin{equation}\label{eq:g_c}
    g_c^2=\omega_A(\omega_C^2+\kappa^2)/(4\lambda_\mathrm{max}\omega_C),
\end{equation}
where $\lambda_\mathrm{max}$ is the largest eigenvalue of the matrix $J_{ij}$ which describes the cavity mediated interactions between atomic clusters $i$ and $j$, and is given by $J_{ij}=\sum_{k=1}^{n_m}g_{ik}g_{jk}$. In the experiments conducted so far, disordered connectivity matrices, $J_{ij}$, have been implemented as they realize associative memories~\cite{marsh_enhancing_2021,marsh_quantum-optical_2024} and spin glass physics~\cite{kroeze_directly_2025,marsh_entanglement_2024,marsh_multimode_2025}, but this phase transition also occurs for ordered connectivity.

One of the simplest possible \textit{ordered} coupling configurations for this system is the nearest-neighbor geometry depicted in Fig.~\ref{fig:Fig1}(b), which is obtained by setting the coupling strengths to be 
\begin{equation}\label{eq:coupling}
    g_{ij}=g(\delta_{i-1,j}+\delta_{i,j}).
\end{equation}

This is the coupling configuration that we focus our considerations on for the majority of this work. For simplicity, we also assume perfectly degenerate cavity modes and identical atomic clusters. 
\subsection{Atom-Only Description}\label{ssec:atom_only}
In order to be able to numerically describe the system across the superradiant transition~\cite{marsh_entanglement_2024,buchhold_dicke-model_2013}, despite the presence of multiple cavity modes, we eliminate the cavity modes and move to an atom-only description. We use the method of Jäger et al.~\cite{jager_lindblad_2022} to arrive at an atom-only master equation in Lindblad form.\\
To begin, we find the effective fields which are used to approximate the effect of the cavity modes. There is one effective field operator for each cavity mode, and they can also be used in the place of cavity annihilation operators $\hat{a}_k$ to calculate expectation values for quantities of interest pertaining to the cavity modes~\cite{jager_lindblad_2022}. For an arbitrary coupling configuration the effective fields are given by
\begin{equation}\label{eq:eff_mode_general}
    \hat{\alpha}_k=-\sum_{i=1}^{n_c}\frac{g_{ik}}{\sqrt{N}}\left[\alpha_+\hat{S}^x_i+i\alpha_-\hat{S}_i^y\right],
\end{equation}
while for the coupling geometry of Eq.~\eqref{eq:coupling}, this becomes
\begin{equation}\label{eq:eff_mode}
    \hat{\alpha}_k=-\frac{g}{\sqrt{N}}\left[\alpha_+(\hat{S}^x_k+\hat{S}_{k+1}^x)+i\alpha_-(\hat{S}_k^y+\hat{S}_{k+1}^y)\right],
\end{equation}
where we introduce the coefficients
\begin{equation}\label{eq:alpha_pm}
    \alpha_{\pm}=\frac{1}{(\omega_C+\omega_A)-i\kappa}\pm\frac{1}{(\omega_C-\omega_A)-i\kappa}.
\end{equation}
With these effective field operators we can then write down the effective Hamiltonian in the atom-only description. In general,
\begin{multline}\label{eq:effective_hamiltonian_general}
    \hat{H}_\mathrm{eff}=\omega_A\sum_{i=1}^{n_c}\hat{S}_i^z-\sum_{i,j=1}^{n_c}\frac{J_{ij}}{N}\left[\alpha_+\hat{S}_j^x\hat{S}_i^x\right.\\\left.+i\alpha_-\hat{S}_j^x\hat{S}_i^y+\mathrm{H.c.}\right],
\end{multline}
 where $J_{ij}$ is again the matrix describing the effective coupling between clusters, while for the specific ordered coupling geometry of Eq.~\eqref{eq:coupling} we can write
\begin{multline}\label{eq:effective_hamiltonian}
    \hat{H}_\mathrm{eff}=\omega_A\sum_{i=1}^{n_c}\hat{S}_i^z-\frac{g^2}{N}\sum_{k=1}^{n_m}\left[(\hat{S}_k^x+\hat{S}_{k+1}^x)\left(\alpha_+(\hat{S}_k^x\right.\right.\\\left.\left.+\hat{S}_{k+1}^x)+i\alpha_-(\hat{S}_k^y+\hat{S}_{k+1}^y)\right)+\mathrm{H.c.}\right].
\end{multline}
Finally, the Lindblad master equation in the atom-only description is
\begin{equation}
    \dot{\hat{\rho}}_\mathrm{sys}=-i[\hat{H}_\mathrm{eff},\hat{\rho}_\mathrm{sys}]+\kappa\sum_{k=1}^{n_m}D[\hat{\alpha}_k]\hat{\rho}_\mathrm{sys}.
\end{equation}
 
From Eq.~\eqref{eq:effective_hamiltonian} we note that the effective coupling geometry, for the atomic clusters only, is a nearest-neighbor geometry. The transformation that is used to move to the atom-only description is most valid when $g\ll\kappa,\omega_C$. 

\section{Steady-State Solutions}\label{sec:steady-state-slns}
The first goal of this work is to identify and characterize the stable steady-state solutions of this model on either side of the superradiant phase transition. To do so, we take the mean-field limit, $N\to\infty$, solve for the steady-state solutions of the resulting equations of motion, and then perform a linearized stability analysis of these solutions.
\subsection{Mean-field Equations}\label{sec:MF_eqns}
From the Lindblad master equation in the atom-only description, Eq.~\eqref{eq:MMC_me}, we can obtain equations of motion for the expectation values of the collective spin operators, $\hat{S}_i^\beta$. Taking the mean-field limit of $N\to\infty$ corresponds to neglecting quantum fluctuations and imposing the factorizations $\langle\hat{S}_i^\beta\hat{S}_j^\gamma\rangle=\langle\hat{S}_i^\beta\rangle\langle\hat{S}_j^\gamma\rangle$. Defining the rescaled expectation values $x_i=\langle\hat{S}^x_i\rangle/N,\ y_i=\langle\hat{S}^y_i\rangle/N$ and $z_i=\langle\hat{S}^z_i\rangle/N$, we obtain the following mean-field equations for the nearest-neighbor coupling configuration:
\begin{widetext}
\begin{subequations}\label{eq:mf_sys}
\begin{equation}\label{eq:mf_x}
    \dot{x}_i=-\omega_Ay_i,
\end{equation}
\begin{align}\label{eq:mf_y}
    \dot{y}_i=\omega_Ax_i&+4g^2V_0z_i\left[(1-\delta_{i,1})(x_{i-1}+x_i)+(1-\delta_{i,n_c})(x_i+x_{i+1})\right]\\&+4g^2V_1z_i\left[(1-\delta_{i,1})(y_{i-1}+y_i)+(1-\delta_{i,n_c})(y_i+y_{i+1})\right], \nonumber
\end{align}
\begin{align}\label{eq:mf_z}
    \dot{z}_i=
         &-4g^2V_0y_i\left[(1-\delta_{i,1})(x_{i-1}+x_i)+(1-\delta_{i,n_c})(x_i+x_{i+1})\right]\\&-4g^2V_1y_i\left[(1-\delta_{i,1})(y_{i-1}+y_i)+(1-\delta_{i,n_c})(y_i+y_{i+1})\right], \nonumber
\end{align}
\end{subequations}
\end{widetext}
where we define the coefficients $V_0=\Re(\alpha_+)$ and $V_1=-(\kappa \Re(\alpha_+^*\alpha_-)+\Im(\alpha_-))/2$. For each atomic cluster there is also a spin conservation constraint,
\begin{equation}\label{eq:spin_conservation}
    x_i^2+y_i^2+z_i^2=\frac{1}{4}.
\end{equation}
Together, Eqs.~\eqref{eq:mf_x}-\eqref{eq:mf_z} and~\eqref{eq:spin_conservation} are solved numerically to find the steady-state solutions. We focus our numerical calculations on the set of parameters $\{\omega_A,\kappa\}=\{0.1,1\}\omega_C$, where the atomic dynamics are much slower than the cavity, justifying the transition to an atom-only picture~\cite{jager_lindblad_2022}. While we consider values of $g$ up to $0.7\omega_C$ in these mean-field calculations, we validate these results with alternative numerical methods that do not rely on the atom-only approximation in Sec.~\ref{sec:finite-size-clusters}. In addition, the atom-only approximation has already been seen to be valid across the superradiant phase transition in the standard Dicke model for the same choices of system parameters that we use~\cite{jager_lindblad_2022}, providing further confidence in the results that follow.

\begin{figure*}[!ht]
  \includegraphics[width=0.95\textwidth]{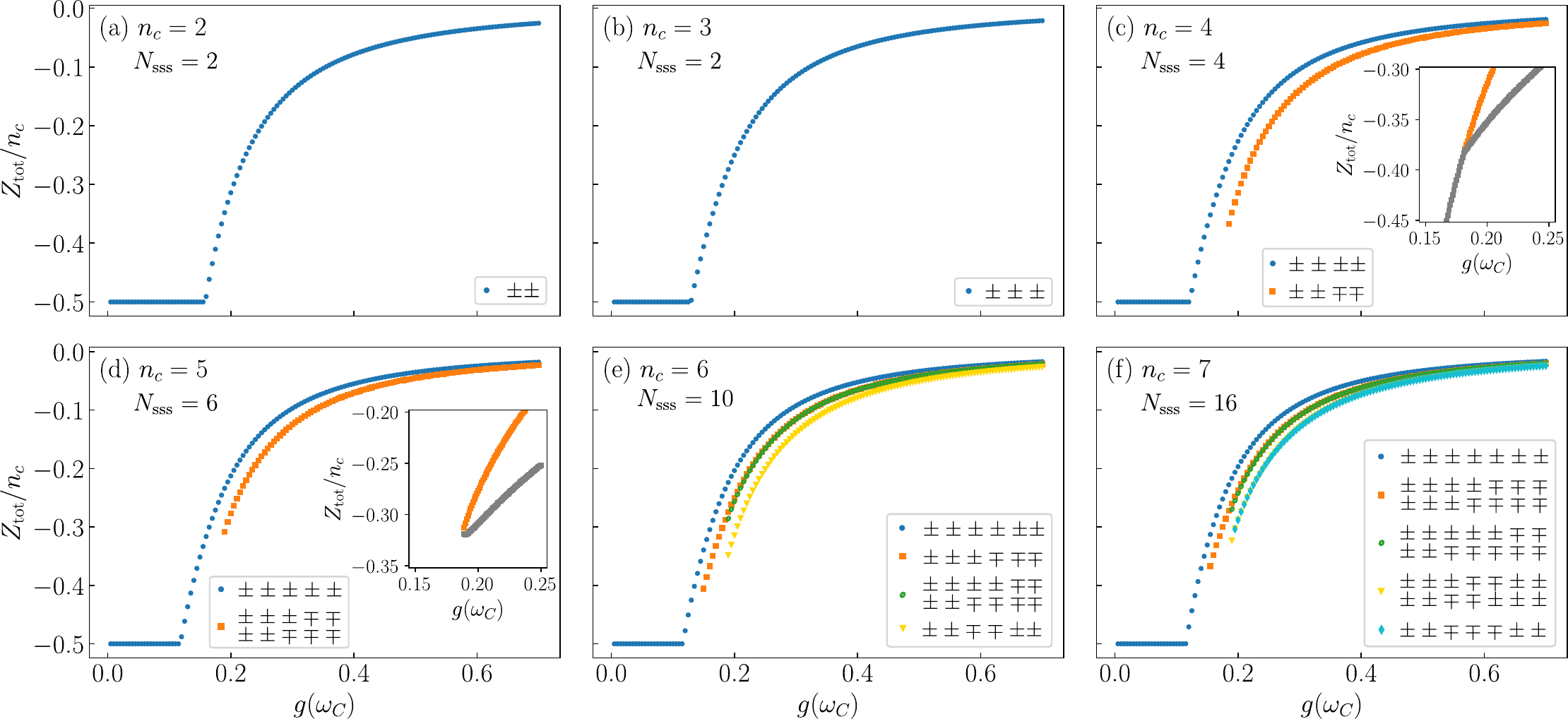}
  \caption{Stable steady-state solutions shown through the total $z$-spin components scaled by the number of clusters plotted as a function of the coupling strength, $g$, for different values of $n_c$. As the number of clusters in the system is increased, the number of stable steady-state solutions in the superradiant phase increases, as given by Eq.~\eqref{eq:fibonacci}. The legends label the pairs of solutions based on the sign of the $x$-spin component of each cluster. Note that some values of $Z_\mathrm{tot}$ correspond to multiple distinct solution pairs due to reflection symmetry about the central cluster. In addition, some of the solutions pairs that are not linked via symmetries, but have the same amount of groups of clusters, also have very similar values of $Z_\mathrm{tot}$ causing the points plotted in this figure to overlap. The insets in (c) and (d) show examples of the bifurcation mechanisms observed so far through which the additional stable steady-state solutions become stable. The gray lines in these insets depict unstable steady-state solutions. The system parameters for these calculations are $\{\omega_A,\kappa\}=\{0.1,1\}\omega_C$.}
  \label{fig:SSS_N}
\end{figure*}

\subsection{Stability Analysis}\label{sec:stability_anlaysis}
In order to determine whether a steady-state solution is stable, we make use of the Jacobian of the system of differential equations (Eqs.~\eqref{eq:mf_sys}). When evaluated at a given steady-state solution, the Jacobian provides a linearization about that point and its eigenvalues, $\lambda_i$, determine the stability of the steady-state solutions~\cite{strogatz_nonlinear_2018}. Any steady-state solution for which there exists a $\lambda_i$ with $\Re(\lambda_i)>0$ is classified as unstable while steady-state solutions for which $\Re(\lambda_i)<0$ for all $\lambda_i$ are stable. Since the system of differential equations of interest is non-linear, care must be taken when assessing the stability of steady-state solutions which have a $\lambda_i$ with $\Re(\lambda_i)=0$.

For the system as it is written in Eqs.~\eqref{eq:mf_sys}, the spin conservation constraint for each cluster (Eq.~\eqref{eq:spin_conservation}) has not been taken into account. As a result, every steady-state solution has one $\lambda_i$ with $\lambda_i=0$. If we were to transform the system variables using a stereographic projection which takes these constraints into account~\cite{stitely_nonlinear_2020}, these eigenvalues would not be present, reassuring us that these eigenvalues with do not affect the stability of the steady-state solutions. The effects of additional purely imaginary eigenvalues will be discussed in more detail in Secs.~\ref{sec:persistent_oscillations_SR} and~\ref{sec:persistent_oscillations_NP}, where we show that in this coupling geometry these steady-state solutions are also stable.

To be able to compare across systems containing different numbers of clusters, we define the total $z$-spin component scaled by the number of clusters in the system, $Z_\mathrm{tot}/n_c=\sum_{i=1}^{n_c}z_i/n_c$. In Fig.~\ref{fig:SSS_N} we plot this quantity for all steady-state solutions which are stable, i.e., for which $\Re(\lambda_i)\leq0$ for all $\lambda_i$, for systems containing between two and seven clusters, as the value of the coupling constant, $g$, is varied. As predicted in~\cite{marsh_entanglement_2024} the nature and number of the stable steady-state solutions changes at the critical coupling strength $g_c$ (given by Eq.~\eqref{eq:g_c}). As in the standard dissipative Dicke model (see for example~\cite{dimer_proposed_2007,kirton_introduction_2019}), the stable steady-state solutions for $g<g_c$, are given by $z_i=-1/2$, with $x_i=y_i=0~\forall i$. From the expressions for the effective fields, Eq.~\eqref{eq:eff_mode}, we see that the cavity modes are unoccupied. This is known as the normal phase. For $g>g_c$, the stable steady-state solutions satisfy $x_i\neq0$, and occur in pairs, linked by the transformation $(x_i)\to-(x_i)\forall i$, as in the standard Dicke model, due to the symmetry associated with the conservation of the parity of the number of excitations~\cite{kirton_introduction_2019}. As $x_i\neq0$, Eq.~\eqref{eq:eff_mode} indicates that the cavity modes no longer have to be empty, as expected in the superradiant phase.

For geometries with more than three clusters, we observe additional stable steady-state solutions in the superradiant phase. As seen in Fig.~\ref{fig:SSS_N}, the number of these additional stable steady-state solutions in the superradiant phase increases as the number of clusters, $n_c$, in the system increases. These solutions can be labeled by the signs of the $x$-spin component of the clusters. For a given solution we can write down the label
\begin{equation}
    \bm{\sigma} = (\sigma_{1}, \dots, \sigma_{n_{c}}), \quad \sigma_{i} \in \{\pm 1\},
\end{equation}
where $\sigma_i=\sgn(x_{i})$,
and while the magnitudes of the spin components change as $g$ is varied, this labeling stays the same~\footnote{This labeling only uniquely identifies \textit{stable} steady-state solutions, as there can be multiple additional unstable steady-state solutions with the same sign pattern but different magnitudes of the $x$-spin components.}. Starting from this labeling, and grouping pairs of steady-state solutions linked by the transformation $(x_i)\to-(x_i)\forall i$ together, we can count the number of clusters in each group that has the same $x$-spin component sign, e.g., $\pm\pm\pm\mp\mp\to\{3,2\}$. We then identify the pairs of stable steady-state solutions in the superradiant phase for $n_c$ clusters as all the possible compositions of $n_c$ which do not include 1, where compositions of an integer $n$ refer to ways of writing $n$ as the sum of a sequence of positive integers. We have observed this pattern for systems with up to eight
clusters through numerical calculations. Through combinatorial arguments~\cite{stanley_enumerative_1997,hoggatt_jr_compositions_1969}, we can express the total number of observed stable steady-state solutions in the superradiant phase, $N_\mathrm{sss}$, in terms of the Fibonacci numbers ($F_n$),
\begin{equation}\label{eq:fibonacci}
    N_\mathrm{sss}(n_c)=2F_{n_c-1}.
\end{equation}

We prove below that stable steady-state solutions following this observed pattern can be found for all $n_c$. This results in an exponential growth of $N_\mathrm{sss}$, which we can characterize for $n_c \gg 1$ as
\begin{equation}\label{eq:fib_exp}
    N_\mathrm{sss}\gtrsim(2/\sqrt{5})\varphi^{n_c-1},
\end{equation}
where $\varphi=(1+\sqrt{5})/2$ is the golden ratio and we use the inequality since our proof does not rule out the possibility of additional steady states.

For the proof we describe the sign patterns corresponding to stable steady state solutions via a counting variable,
\begin{equation}
    q_{i} = \bm{1}_{\{i>1 \text{ and } \sigma_{i-1} = \sigma_{i}\}} + \bm{1}_{\{i<n_{c} \text{ and } \sigma_{i+1} = \sigma_{i}\}},
\end{equation}
which for a given cluster $i$, with sign $\sigma_{i}$, counts how many nearest-neighbors of cluster $i$ have the same sign. Defining a singleton as a cluster $i$ with $q_i=0$ allows us to say that all the stable steady-state solutions correspond to no-singleton solutions. For example, the pattern $\pm\pm\mp\pm\pm$, which does not correspond to a stable steady-state, has a singleton $\mp$ at the third cluster, whereas $\pm\pm\pm\mp\mp$ has no singletons and corresponds to a stable steady-state as seen in Fig.~\ref{fig:SSS_N}(d). In Fig.~\ref{fig:SSS_N} we also observe that for all the stable steady-state solutions we have $Z_{tot}/n_c\to0$ as $g$ increases, corresponding to $x_i\to\sigma_i/2$. In order to provide a justification that the Fibonacci scaling, Eq.~\eqref{eq:fibonacci}, which was numerically observed for small $n_c$, holds for arbitrary $n_c$, we define the dimensionless parameter
\begin{equation}\label{eq:epsilon}
    \epsilon=\frac{\omega_A}{g^2V_0}\propto \left(\frac{g_c}{g}\right)^2 ,
\end{equation}
and consider a perturbative expansion of these no-singleton solutions with $x_i=\sigma_i/2$ and $z_i=0$, about $\epsilon=0$. The details of this perturbative stability analysis can be found in Appendix~\ref{app:Perturbative_analysis}, but we summarize the main results here. Firstly we find that there is an analytic continuation of these solutions for small $\epsilon$, justifying the perturbative expansion given by
\begin{equation}
    z_{i}(\epsilon) = -\dfrac{1}{8 q_{i}} \epsilon + \mathcal{O}(\epsilon^{3}),\ \ \ \ \ 
    x_{i}(\epsilon) = \dfrac{\sigma_{i}}{2} - \dfrac{\sigma_{i}}{64 q_{i}^{2}} \epsilon^{2} + \mathcal{O}(\epsilon^{4}).
\end{equation}
We note that if $q_i=0$ for any $i$, i.e., if the solution contains a singleton, this perturbative expansion fails.

Following this, we apply Lyapunov stability theory~\cite{khalil_nonlinear_2002}, to the mean-field equations, Eqs.~\eqref{eq:mf_sys}, linearized about these no-singleton solutions, to conclude that the no-singleton solutions are Lyapunov stable for sufficiently small $\epsilon$. In doing so, we discover that the dynamics of the fluctuations can be expressed as a generalized damped harmonic oscillator, with a damping term proportional to $V_1$, such that $V_1>0$ ensures positive damping, implying stability of these no-singleton solutions. This argument applies for all values of $n_c$, showing that the pattern in the stable steady-state solutions that was numerically observed for small $n_c$ holds for all $n_c$ in the limit that $\epsilon\to0$. An alternative proof for the existence of the no-singleton steady-state solutions, in terms of a fixed point argument that does not require a perturbative expansion, is presented in Appendix~\ref{sec:existence}. This proof provides us with a bound of $\epsilon\leq3/4$ for which we can guarantee the existence, and stability, of all the no-singleton stable steady-state solutions, regardless of the number of clusters in the system~\footnote{We note that this analysis does not rule out the possibility of additional stable steady-state solutions, which do not follow this pattern.}. From Eq.~\eqref{eq:epsilon}, we can turn this requirement for $\epsilon$ into a requirement for $g$, showing that all the no-singleton stable steady-state solutions exist once the system is sufficiently far into the superradiant regime. For the bound of $\epsilon\leq3/4$, and the parameters $\{\omega_A,\kappa\}=\{0.1,1\}\omega_C$ used throughout this paper, we obtain $g\geq0.37$, which matches the numerical results presented in Fig.~\ref{fig:SSS_N} as no additional stable steady-state solutions appear beyond this value of $g$. We note that the bound of $\epsilon\leq3/4$ is likely a pessimistic bound and could possibly be improved via the optimization of constants.

\subsection{Persistent Oscillations in the Superradiant Phase}\label{sec:persistent_oscillations_SR}
We now return to the steady-state solutions which have additional eigenvalues of the Jacobian with $\Re(\lambda_i)=0$ to fully determine their stability. The solutions which this applies to are the normal phase steady-state solutions with $z_i=-1/2$ and $x_i=y_i=0~\forall i$, and the steady-state solutions in the superradiant phase in which clusters are grouped into superradiant pairs with alternating signs
of the $x$-spin component, e.g., $\pm\pm\mp\mp\pm\pm$.

\begin{figure}[t]
  \includegraphics[width=0.9\linewidth]{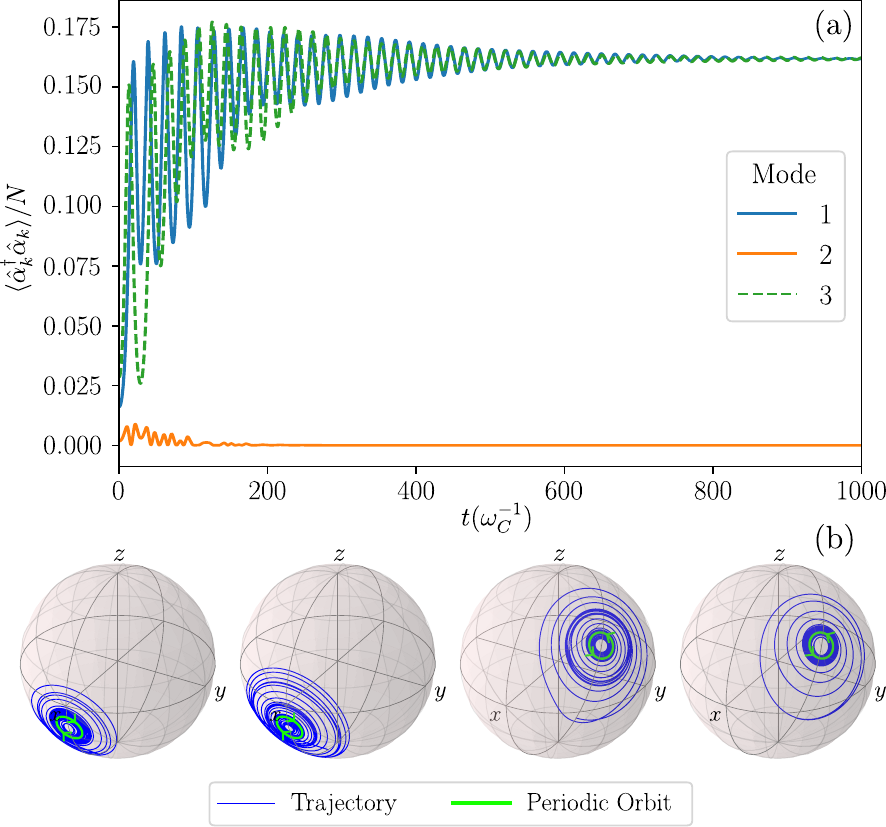}
  \caption{Example mean-field evolution of a four cluster system towards an oscillatory steady-state solution in the superradiant phase. (a) The cavity field occupations converge to fixed values, with finite occupation in the cavity modes that connect clusters within superradiant pairs (modes 1 and 3), and with zero occupation in the cavity mode connecting two superradiant pairs (mode 2). (b) The paths traced out by the spin of each cluster on its own generalized Bloch sphere (blue) converge to closed phase space orbits (green), leading to oscillations in the spin components. The orbits for the first two clusters are identical and the orbits for the last two clusters are identical, and the two pairs are connected by the transformation $x\to-x$. The initial conditions used to generate these plots are $(x_i,0,-\sqrt{1/4-x_i^2})$ with $x_1=0.2,\ x_2=0.1,\ x_3=-0.2$ and $x_4=-0.2$. The parameters were $\{\omega_A,\kappa,g\}=\{0.1,1,0.3\}\omega_C$.}
  \label{fig:4_SR_oscillations}
\end{figure}

We begin by considering the oscillatory steady-state solutions in the superradiant phase. The structure of the superradiant steady-state oscillatory solutions is such that the clusters all share the same magnitude of the $x$-spin component, so that due to the alternating signs of the $x$-spin components, every second cavity mode is completely empty, as seen in Eq.~\eqref{eq:eff_mode}. This indicates that the different pairs of clusters effectively decouple. This is shown in Fig.~\ref{fig:4_SR_oscillations}(a) where after an initial evolution away from the initial conditions, the occupation of the second cavity mode quickly becomes zero. Because of this decoupling, we can understand the oscillatory nature of the solutions by looking at the simpler case of two atomic clusters coupled equally to a single cavity mode. In this case, the Hamiltonian,
\begin{equation}\label{eq:2_clusters}
    \hat{H}=\omega_A(\hat{S}_1^z+\hat{S}_2^z)+\omega_C\hat{a}^\dagger\hat{a}+\frac{2g}{\sqrt{N}}(\hat{a}+\hat{a}^\dagger)(\hat{S}^x_1+\hat{S}^x_2),
\end{equation}
can be identified with that of the standard Dicke model describing a single cluster containing $2N$ atoms coupled to a single cavity mode, by defining new collective spin operators, $\hat{\tilde{S}}^{\beta}=\hat{S}^\beta_1+\hat{S}^\beta_2$ ($\beta=x,~y,~z$). In contrast to the full multimode model, the total spin, $S^2=(x_1+x_2)^2+(y_1+y_2)^2+(z_1+z_2)^2$, is now an additional conserved quantity of the system making the non-linear system of differential equations, Eq.~\eqref{eq:mf_sys}, a conservative system, which allows us to conclude that this steady-state solution which has an additional eigenvalue of the Jacobian with $\Re(\lambda_i)=0$ is a stable center~\cite{strogatz_nonlinear_2018}. This means that this steady-state solution corresponds to an infinite number of closed phase space orbits centered around the mean-field solution, leading to oscillatory solutions.

We do not see these persistent oscillations in the standard Dicke model because in modeling the atoms (or clusters of atoms in this case) collectively we make the assumption that they are indistinguishable and thus impose permutational invariance on the states of the components of the cluster. When considering two clusters coupled to a single cavity mode this requirement of permutational invariance is partially lifted and we can choose to initialize the two clusters in different states. As a result, oscillatory behavior is only observed when the system is initialized in a inhomogeneous initial state~\cite{iemini_dynamics_2024}. We can see this from expressions for these oscillatory solutions. The derivation of these expressions is presented in Appendix~\ref{app:Osc_sols}.
\begin{subequations}\label{eq:osc_sols_SR}
\begin{align}
    x_{\{1,2\}}(t)&=\frac{1}{2}\left(x_\mathrm{tot}^\mathrm{ss}\pm\frac{\omega_AA}{4V_0S}\cos(4V_0St+\phi)\right),\\
    y_{\{1,2\}}(t)&=\pm\frac{A}{2}\sin(4V_0St+\phi),\\
    z_{\{1,2\}}(t)&=\frac{1}{2}\left(z_\mathrm{tot}^\mathrm{ss}\pm\frac{x_\mathrm{tot}^\mathrm{ss}A}{S}\cos(4V_0St+\phi)\right),
\end{align}
\end{subequations}
where $x_\mathrm{tot}^\mathrm{ss}=\pm\sqrt{S^2-\omega_A^2/16V_0^2}$ and $z_\mathrm{tot}^\mathrm{ss}=-\omega_A/4V_0$ are the steady-state solutions for $x_\mathrm{tot}=x_1+x_2$ and $z_\mathrm{tot}=z_1+z_2$ respectively, and $\phi$ is a phase that accounts for the effect of the initial dynamics. We have introduced $A$, which is related to the total spin, $S$, by $A=\sqrt{1-S^2}$, and is a measure of the permutational symmetry of the initial state~\cite{iemini_dynamics_2024}. For a permutationally symmetric state where $x_1=x_2$, $y_1=y_2$ and $z_1=z_2$, we have that $S=1$ and $A=0$ and thus the oscillatory components of the solutions vanish. Equations~\eqref{eq:osc_sols_SR} also show how strongly the oscillatory solutions depend on the initial conditions, in particular on the total spin, which determines the center, amplitude and frequency of the particular oscillatory solution the system will converge to.

In the case where we have an even value of $n_c$ greater than two, the total spin is no longer a conserved quantity. However, in the subspace spanned by 
\begin{subequations}\label{eq:subspaceSR}
\begin{equation}
    c_x-x_{4k-3}=x_{4k-2}=-x_{4k-1}=c_x+x_{4k},
\end{equation}
\begin{equation}
    y_{4k-3}=-y_{4k-2}=y_{4k-1}=-y_{4k},
\end{equation}
\begin{equation}
    z_\mathrm{tot}^\mathrm{ss}-z_{4k-3}=z_{4k-2}=z_{4k-1}=z_\mathrm{tot}^\mathrm{ss}-z_{4k},
\end{equation}
\end{subequations}
where $k=1,~2,~3,\dots$, $z_\mathrm{tot}^\mathrm{ss}=-\omega_A/4V_0$ and $c_x$ is an arbitrary constant, the combined spin of the spins in each of the superradiant pairs,
\begin{equation}\label{eq:spin_pairs}
    S_m^2=(x_{2m-1}+x_{2m})^2+(y_{2m-1}+y_{2m})^2+(z_{2m-1}+z_{2m})^2, 
\end{equation}
with $m=1,~2,~3,\dots$, is conserved. The superradiant oscillatory steady-state solutions are clearly contained within this subspace, and it is simple to show that this subspace remains closed under evolution of the mean-field equations. Within this subspace the conserved combined spins of the superradiant spin pairs act to stabilize the oscillatory solutions, and we can generalize the expressions in Eqs.~\eqref{eq:osc_sols_SR} by following the pattern given by the requirements of the subspace. An example of this can be seen in Fig.~\ref{fig:4_SR_oscillations}. Note that in the case of more than two clusters, the combined spins are not conserved until the system enters the subspace described by Eqs.~\eqref{eq:subspaceSR}, so the parameters characterizing the oscillations are difficult to predict for arbitrary initial conditions. 

An alternative justification for the presence of the persistent oscillations for this set of superradiant solutions comes from the perturbative stability analysis introduced in the previous section, and discussed in detail in Appendix~\ref{sec:perturbation_oscillations}. Considering the asymptotic stability of the no-singleton solutions reveals that the only no-singleton solutions which are not asymptotically stable, and for which trajectories do not converge to the steady-state solution as $t\to\infty$, are those for which  $q_i = 1$ for all $i$, i.e., those for which the sign pattern consists of alternating blocks of length two. These correspond precisely to the solutions identified above which display persistent oscillations. In particular, in the damped harmonic oscillator model for the fluctuations, these solutions do not experience any damping, providing an intuition for the observed persistent oscillations.

\subsection{Persistent Oscillations in the Normal Phase}\label{sec:persistent_oscillations_NP}
\begin{figure}[t]
  \includegraphics[width=0.9\linewidth]{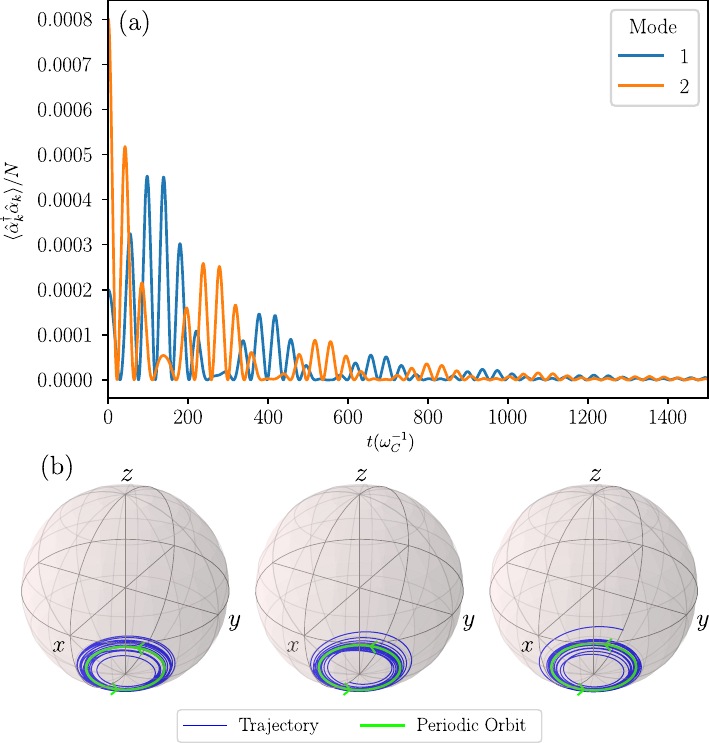}
  \caption{Example mean-field evolution of a three cluster system towards an oscillatory steady-state solution in the normal phase. (a) The cavity field occupations for both modes converge to zero as expected for the normal phase. (b) The paths traced out by the spin of each cluster on its own generalized Bloch sphere (blue) converge to closed phase space orbits (green), leading to oscillations in the spin components. The orbits for all three clusters trace out the same circle, but the orbit of the middle cluster is out of phase with that of the two edge clusters so that $x_i=-x_{i+1}$ and $y_i=-y_{i+1}$. The initial conditions used to generate these plots are $(x_i,0,-\sqrt{1/4-x_i^2})$ with $x_1=-0.2,\ x_2=0.1,$ and $x_3=-0.3$. The parameters were $\{\omega_A,\kappa,g\}=\{0.1,1,0.1\}\omega_C$.}
  \label{fig:3_NP_oscillations}
\end{figure}

In the normal phase, the only solution which is not unstable corresponds to an oscillatory solution. In the case of two clusters coupled identically to a single cavity mode, the same argument as in the superradiant phase, in terms of spin conservation, can be applied in the normal phase. Thus the steady-state solution is again a stable center corresponding to an infinite number of oscillatory solutions centered around the mean-field solution. In the normal phase the oscillations are centered at $x_i=y_i=0~\forall i$ with the $z$-spin component remaining constant at a value determined by the initial conditions. The $x$ and $y$-spin component oscillations in adjacent clusters occur out of phase so that $x_1=-x_2$ and $y_1=-y_2$. This is summarized by the following equations (for which the derivation is also presented in Appendix~\ref{app:Osc_sols}),
\begin{subequations}\label{eq:osc_sols_NP}
    \begin{align}
    x_{\{1,2\}}(t)&=\pm\frac{A}{2}\cos(\omega_At+\phi), \\
    y_{\{1,2\}}(t)&=\pm\frac{A}{2}\sin(\omega_At+\phi),\\
    z_{\{1,2\}}(t)&=-\frac{S}{2},
\end{align}
\end{subequations}
where $A$, $S$ and $\phi$ are as in the superradiant phase. As seen in Eq.~\eqref{eq:eff_mode}, and Fig.~\ref{fig:3_NP_oscillations}, the fact that $x_1=-x_2$ and $y_1=-y_2$ results in an empty cavity mode even though the spin clusters are not completely spin down.

When there are more than two clusters in the system the total spin is again no longer a conserved quantity. However, as in the case of the superradiant oscillations we can identify a subspace that remains closed under evolution of the mean-field equations that contains the oscillatory solutions and within which a conserved quantity exists to stabilize the persistent oscillations. In the normal phase, this subspace is the one spanned by
\begin{align}
    x_{k+1}=-x_k, && y_{k+1}=-y_k, && z_{k+1}=z_k,
\end{align}
where $k=1,~2,~3\dots$ and the combined spins of adjacent spin clusters, $S^2_m$, as in Eq.~\eqref{eq:spin_pairs}, are again conserved quantities. In addition, the energy scaled by the number of atoms per cluster, $E=\langle\hat{H}_\mathrm{eff}\rangle/N$, that is
\begin{multline}
    E=\omega_A\sum_{i=1}^{n_c}z_i-g^2\sum_{k=1}^{n_m}\left[(x_k+x_{k+1})\left(\alpha_+(x_k+x_{k+1})\right.\right.\\\left.\left.+i\alpha_-(y_k+y_{k+1})\right)+\mathrm{c.c.}\right],
\end{multline}
is also conserved in this subspace, and thus along these periodic orbits. Because we are including dissipation via the cavity modes in our model, it is unexpected that there should be energy conserving trajectories. However, due to a careful balance of the processes that exchange energy between the various components of the system, and in this case the fact that no energy is lost via dissipation because the cavity modes stay empty along the trajectories, such trajectories are possible, and similar energy conserving, periodic orbits have also been observed in other variants of the open Dicke model~\cite{stitely_nonlinear_2020,adiv_nonlinear_2026}. 

The expressions given by Eqs.~\eqref{eq:osc_sols_NP} can, as in the superradiant phase, be generalized to the case of multiple spin clusters by following the pattern of the subspace, noting again that the amplitude of the oscillations reached from a particular initial condition is difficult to predict. An example trajectory for three clusters is shown in Fig.~\ref{fig:3_NP_oscillations}.

\section{Finite-size Clusters}\label{sec:finite-size-clusters}
All the results presented in the previous sections were obtained in the mean-field limit, where $N\to\infty$ and quantum fluctuations were neglected. From an experimental perspective, this is not realistic and we now consider the effect of quantum fluctuations on the properties of the stable steady-state solutions identified in the mean-field limit in the previous section. To perform this check of the robustness of the phenomenology of the previous section, we make use of the truncated Wigner approximation (TWA). As we do not use an atom-only approximation to eliminate the cavity modes, the comparison of results obtained with the TWA to the mean-field results of the previous section will also allow us to validate the atom-only approximation made when deriving the mean-field equations.

\subsection{Truncated Wigner Approximation}
The truncated Wigner approximation is a semi-classical phase-space method that uses a stochastic ensemble of classical trajectories to enable an approximate treatment of quantum dynamics beyond mean-field theory~\cite{polkovnikov_phase_2010}. The TWA takes into account first order quantum fluctuations and is computationally efficient for finite times, even for many-body systems~\cite{hosseinabadi_user-friendly_2025}.

The Wigner phase-space distribution is related to the density matrix through the Weyl transformation. For a system made up of bosonic modes this is 
\begin{equation}\label{eq:Weyl}
    W_{\hat{\rho}}({\vec{q}},{\vec{p}})=\int d\vec{\eta}\langle\vec{q}-\frac{\vec{\eta}}{2}|\hat{\rho}|\vec{q}+\frac{\vec{\eta}}{2}\rangle e^{i\vec{p}\vec{\eta}/\hbar},
\end{equation} 
where ${\vec{q}}$ and ${\vec{p}}$ are position and momentum phase-space variables~\cite{moyal_quantum_1949}. Applying the Weyl transform to the Lindblad master equation gives an equation for the time evolution of the Wigner distribution~\cite{polkovnikov_phase_2010}. When applying the TWA to the multimode Dicke model, we have chosen to work with the full model rather than the atom-only description, as the TWA is numerically more tractable. To deal with the spin degrees of freedom of the atomic clusters, we make use of the Schwinger representation~\cite{schwinger_angular_1952} before transforming back into a spin representation after the Weyl transformation, where we make use of Bopp operators to deal with product terms~\cite{polkovnikov_phase_2010,kubo_wigner_1964}. Truncating the differential equation for the time evolution of the Wigner distribution at second order (equivalent to discarding terms of order $\hbar^2$ or higher) leads to the following Fokker-Planck equation~\cite{polkovnikov_phase_2010},
\begin{figure*}[t!]
  \includegraphics[width=\textwidth]{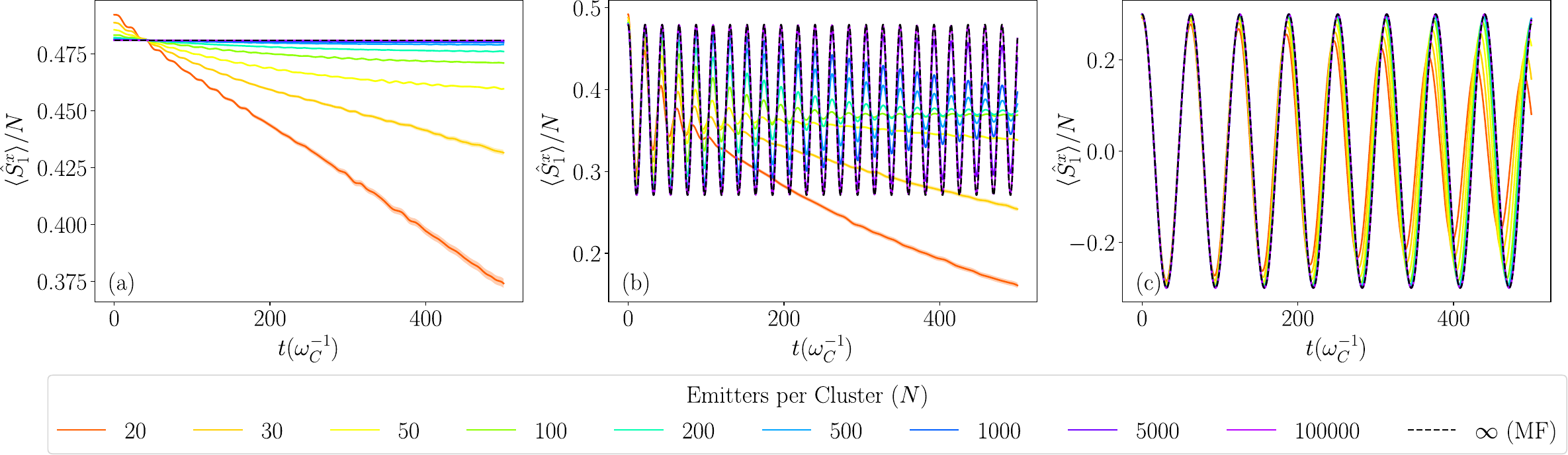}
  \caption{Comparison of the mean-field solutions (dashed black curve) with results obtained using TWA for finite-size clusters with varying numbers of emitters per cluster. The comparison is made for the expectation value of the $x$-spin component of the first cluster scaled by the cluster size $N$. (a) Non-oscillatory superradiant steady-state solution, $\pm\pm\pm\mp\mp$, for five clusters. For large clusters the deviations from the mean-field solution are small indicating stability of the non-oscillatory solution even for finite-size clusters.
  (b) Superradiant phase oscillations for two clusters.
  (c) Normal phase oscillations for two clusters. 
  Oscillations are observed in both phases even for finite-size clusters. These oscillations decay, with more significant decay observed for smaller clusters.
  The parameters used were $\{\omega_A,\kappa\}=\{0.1,1\}\omega_C$ and $g=0.1\omega_C$ for the normal phase in (c), and $g=0.3\omega_C$ for the superradiant phase in (a) and (b). For the oscillatory solutions the specific mean-field periodic orbit is given by $A=0.6$ and $\phi=0$ in Eqs.~\eqref{eq:osc_sols_SR} and~\eqref{eq:osc_sols_NP} for the superradiant phase and the normal phase respectively. For $N\leq100$ there is a significant deviation between the initial expectation values and the mean-field solutions. This is due to the $\mathcal{O}(1/N)$ term in Eq.~\eqref{eq:dist_spin} which is neglected when carrying out the sampling for the initial spin distribution. We show $\pm2$ standard errors around the mean by the shaded region around the curves. This is only visible for some curves in (a) and (b) as for all other curves the error band is thinner than the curve itself.
  }
  \label{fig:TWA_N}
\end{figure*}
\begin{widetext}
\begin{multline}\label{eq:FP}
    \frac{\partial W_{\hat{\rho}}}{\partial t}=\sum_{i=1}^{n_c} \left[\frac{\partial}{\partial X_i}\left(\omega_AY_i\right)+\frac{\partial}{\partial Y_i}\left(-\omega_AX_i+2\sum_{j=1}^{n_m}2g_{ij}u_jZ_i\right)+\frac{\partial}{\partial Z_i}\left(-2\sum_{j=1}^{n_m}2g_{ij}u_jY_i\right)\right]W_{\hat{\rho}}
    \\+\sum_{j=1}^{n_m}\left[\frac{\partial}{\partial u_j}\left(-\omega_Cv_j+\kappa u_j\right)+\frac{1}{2}\frac{\partial^2}{\partial u_j^2}\left(\frac{\kappa}{2N}\right)+\frac{\partial}{\partial v_j}\left(\sum_{i=1}^{n_c}2g_{ij}X_i+\omega_Cu_j+\kappa v_j\right)+\frac{1}{2}\frac{\partial^2}{\partial v_j^2}\left(\frac{\kappa}{2N}\right)\right]W_{\hat{\rho}},
\end{multline} 
\end{widetext}
where $X_i=S^x_i/N,~Y_i=S_i^y/N$ and $Z_i=S^z_i/N$ are scaled versions of the spin phase-space variables, and $u_j=q_j\sqrt{\omega_C/2N}$ and $v_j=p_j/\sqrt{2N\omega_C}$ are scaled versions of the position and momentum phase-space variables of the cavity modes respectively. The Fokker-Planck equation, Eq.~\eqref{eq:FP}, can then be mapped to the following set of stochastic (Itô) differential equations~\cite{gardiner_stochastic_2010},
\begin{subequations}
\begin{align}
    \frac{dX_i}{dt}&=-\omega_AY_i,\\
   \frac{dY_i}{dt}&=\omega_AX_i-2\sum_{j=1}^{n_m}2g_{ij}u_jZ_i,\\
     \frac{dZ_i}{dt}&=2\sum_{j=1}^{n_m}2g_{ij}u_jY_i,\\
    du_j&=(\omega_Cv_j-\kappa u_j)dt+\sqrt{\frac{\kappa}{2N}}dW_{u_j},\\
   dv_j&=-\left(\sum_{i=1}^{n_c}2g_{ij}X_i+\omega_Cu_j+\kappa v_j\right)dt+\sqrt{\frac{\kappa}{2N}}dW_{v_j}.
\end{align}
\end{subequations}
To obtain expectation values for the phase-space variables, which can then be compared to the expectation values obtained from the mean-field treatment, we numerically evolve the stochastic differential equations for the phase-space variables, for many different initial conditions that are sampled from the Wigner distribution of the initial state, and then average values of the desired phase-space variable over the different trajectories. Quantum corrections appear through the noise that is inherent to stochastic differential equations and through the finite width of the initial Wigner distribution. In the multimode Dicke model, both of these effects become less significant as $N\to\infty$, and we expect the approximation to become accurate in this limit~\cite{huber_phase-space_2021}.

To sample the initial Wigner distribution, we assume a product state of coherent states for both the cavity modes and the spin clusters, centered on the mean-field solutions we are comparing to. Thus, the Wigner distributions for each component can be sampled independently. For the cavity modes this results in Gaussian distributions for the position and momentum phase-space variables $u_j, v_j$, which are thus sampled from the probability distribution 
\begin{equation}\label{eq:dist_uv}
    P(w)=\sqrt{\frac{2N}{\pi}}e^{-2N(w-w_0)^2}.
\end{equation}
For the spin variables the initial spin-coherent distribution can be obtained by sampling from the known spin-coherent distribution centered at the north pole of the generalized Bloch sphere, and then applying a rotation to the resulting spin state. The Wigner distribution for the spin-coherent state at the north pole of the generalized Bloch sphere is~\cite{klimov_classical_2005}:
\begin{equation}\label{eq:dist_spin}
    W(\phi,\theta)=\cos^{N-1}(\theta)(1+\cos(\theta))+\mathcal{O}(1/N).
\end{equation}
Thus, $\phi$ is sampled from a uniform distribution on $[0,2\pi]$ and $\theta$ is sampled using the distribution given in Eq.~\eqref{eq:dist_spin}. We note that this (Wigner) distribution is not always strictly positive (in the case of even values of $N$) as needed for applying the TWA, however, for large enough $N$, the negative contributions become comparatively small and can safely be neglected. To do so, we set $W(\phi,\theta)=0$ for $\theta>\pi/2$ as the negative contributions only occur on the southern hemisphere of the generalized Bloch sphere, and for large enough $N$ the spin coherent distributions are sufficiently localized that the discarded positive contributions to the Wigner distribution are also negligibly small.

We use the TWA to check the presence of three different features of the stable steady-state solutions observed in the mean-field when quantum fluctuations are taken into account; the persistent oscillations in the normal phase, the persistent oscillations in the superradiant phase and the existence of multiple stable non-oscillatory steady-state solutions in the superradiant phase.  Figure~\ref{fig:TWA_N} shows that all three of these properties of the steady-state solutions can be observed for finite times in the case of finite cluster sizes. When initialized in the additional non-oscillatory stable steady-state solutions observed in the superradiant phase in the mean-field description, the expectation values of the spin components only show small deviations over time from their initial values, in particular for larger cluster sizes. A similar decay away from the mean-field values is seen for all the non-oscillatory steady-state solutions that are stable at a given value of $g$, indicating that there are indeed multiple equally stable steady-state solutions in the superradiant phase. In both the normal and superradiant phases, oscillations in the spin components can be observed for sufficiently large cluster sizes, with the oscillations decaying at rates depending on the cluster sizes. The plots in Fig.~\ref{fig:TWA_N} show that the TWA results tend towards the mean-field solutions for larger cluster sizes. In particular, for experimentally relevant cluster sizes of $10^5$~\cite{kroeze_high_2023,kroeze_directly_2025} 
the difference between the mean-field solution and the TWA calculations is less than $2.5\times10^{-3}$ for $t<500\omega_C^{-1}$. This agreement of the TWA for large cluster sizes, with the mean-field results that were obtained in the atom-only approximation also justifies the use of the atom-only approximation across the phase transition and into the superradiant regime.
\begin{figure}[t!]
  \includegraphics[width=0.95\linewidth]{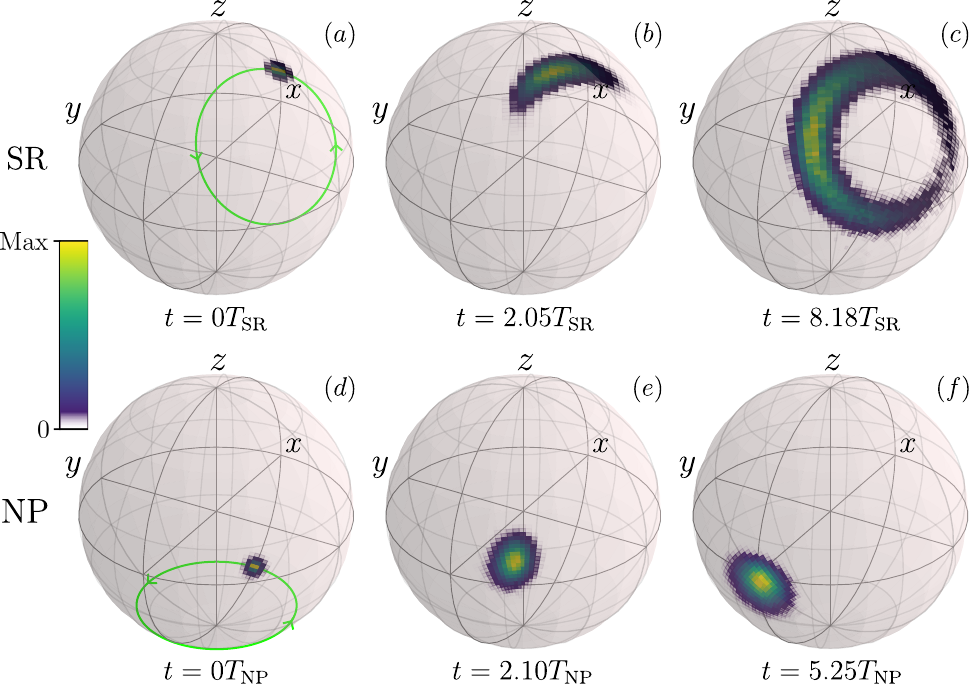}
\caption{Wigner distributions of the first cluster at different times for oscillations in a system of two clusters in the superradiant phase (a-c) and in the normal phase (d-f). In plots (a-c) the period of the superradiant oscillations is given by $T=2\pi/(4V_0S)\simeq22\omega_C^{-1}$ and in plots (d-f) the period of the oscillations in the normal phase is $T=2\pi/\omega_A\simeq63\omega_C^{-1}$ where $N=300$ and the remaining parameter values are $\{\omega_A,\kappa\}=\{0.1,1\}\omega_C$ and $g=0.3\omega_C,~0.1\omega_C$ for the superradiant and normal phase respectively. The initial spin coherent states, centered on the mean-field solutions, are shown in (a) and (d). These are defined by using $A=0.6$ and $\phi=0$ in Eqs.~\eqref{eq:osc_sols_SR} and~\eqref{eq:osc_sols_NP} respectively. In both cases, the Wigner distribution orbits the generalized Bloch sphere, along the mean-field orbits shown in green in (a) and (d), and spreads out over time, but in the superradiant phase this spread is mostly along the orbit, while in the normal phase the spreading is more uniform in all radial directions. Note that the Wigner distributions are plotted as histograms of the individual TWA trajectories, with a a resolution of 100 bins in both angular directions. The color-maps of the distributions are normalized according to their maximum value independently at each value of $t$, and bins with a count value below 10\% of the maximum value fade to transparent for visibility.}
  \label{fig:Wigner_functions}
\end{figure}

To quantify how the size of the clusters impacts the decay away from the mean-field solutions in the TWA calculations, we consider the arc-length $s$ between each point in the Wigner distribution and the point on the generalized Bloch sphere that the mean-field oscillations are centered around. In the case of the solutions corresponding to persistent oscillations, the evolution of the Wigner distribution is depicted in Fig.~\ref{fig:Wigner_functions}. In the normal phase, the entire Wigner distribution orbits around the generalized Bloch sphere, roughly along the mean-field periodic orbit, spreading out in all directions while doing so. In the superradiant phase, the Wigner distribution also orbits around the generalized Bloch sphere roughly along the mean-field periodic orbit, but the distribution preferentially spreads out along this orbit until it forms a closed ring, which then continues to increase in width. Because of the different paths the Wigner distribution takes, we define the arc-length with respect to the center of the orbit in each case, as opposed to using the same reference point for both phases. In particular we investigate how the variance of this arc-length, $\mathrm{Var}(s)$, changes with time and with $N$, which quantifies the spread of the Wigner distribution in the radial direction. In the case of the non-oscillating stable steady-state solution we simply define the arc-length with respect to the mean-field solution, which lies roughly at the center of the distribution.

\begin{figure}[t!]
  \includegraphics[width=\linewidth]{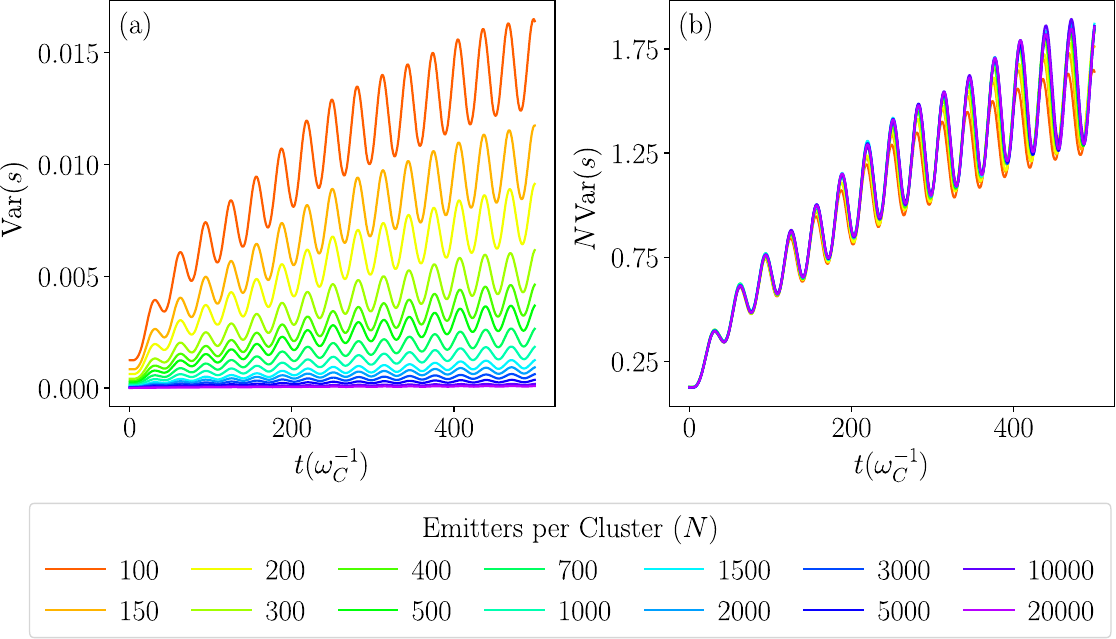}
\caption{(a) Variance of the arc-length of individual trajectories within the TWA evolution, with respect to the south pole, in the first cluster of a two-cluster normal phase oscillation for different cluster sizes. (b) The same arc-length variances as plotted in (a) scaled by the number of atoms per cluster. As the scaled arc-length variances collapse onto a single curve, the dependence of the arc-length variance on $N$ is $1/N$. The system parameters used here are $\{\omega_A,\kappa,g\}=\{0.1,1,0.1\}\omega_C$, and $A=0.6$, $\phi=0$ to determine the particular periodic orbit.}
  \label{fig:Variance}
\end{figure}

While the overall behavior of the arc-length variances are hard to describe with a simple function, we observe that the rate at which the variance in the arc-length increases is inversely proportional to $N$, the number of emitters per cluster, as the variance in the arc-length scales with $1/N$ for all times as seen in Fig.~\ref{fig:Variance} where this is demonstrated for oscillations in the normal phase with $A=0.6$. The same scaling is observed in the superradiant phase as well, both for the oscillatory solutions and the non-oscillatory stable steady-states. In addition, this scaling also remains the same regardless of which solution is considered or, where relevant, which amplitude and phase are chosen to define the oscillations. The fact that the variance in the arc-length, which is a measure of dephasing, has this dependence on $N$ is not particularly surprising, as the variances of the (initial) Wigner distributions for coherent states scale as $1/N$, both for the coherent states of the cavity field given by Eq.~\eqref{eq:dist_uv}, and for the spin coherent states (see for example ~\cite{sanchez-soto_phase_2025}). In addition, the variance of the noise terms in the stochastic differential equations, which also contribute to the dephasing, scales as $1/N$ as well.

In Fig.~\ref{fig:Variance}(b) the curves corresponding to the smallest cluster sizes can be seen to deviate slightly from the other curves, especially at longer times. In the case of the normal phase oscillations, this happens due to the larger Wigner distributions beginning to overlap with the south pole of the generalized Bloch sphere, which, as the center of the oscillations and the stable steady-state solution in the absence of oscillations, will significantly affect the dynamics of the trajectories that get very close to it. The extent to which this is an issue depends on the initial conditions, and how close the mean-field orbit is to the south pole.

\begin{figure*}[t!]
\centering
    \includegraphics[width=\textwidth]{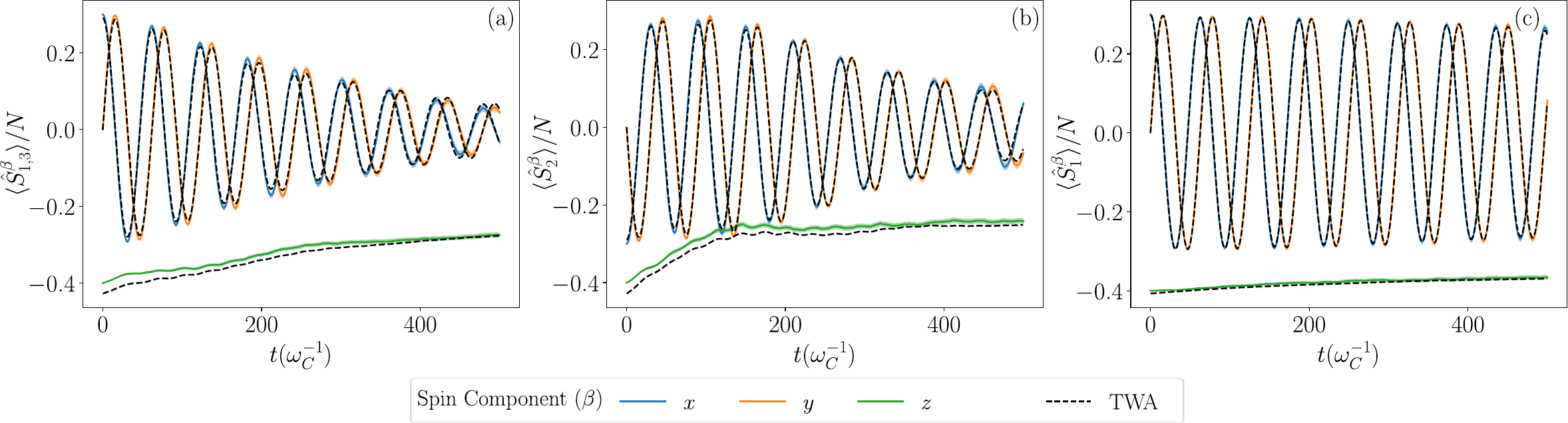}
    \caption{(a-b) Comparison of TWA with HOPS for three clusters for oscillations in the normal phase for the outer and middle spin clusters respectively. The solid blue, orange and green lines show the scaled $x$, $y$, $z$-spin component expectation values obtained using HOPS respectively, while the dashed black curves show the results obtained using the TWA.
    These calculations were done for $N=14$ atoms per cluster, $\{\omega_A,\kappa,g\}=\{0.1,1,0.1\}\omega_C$, and $A=0.6$, $\phi=0$.
    The disagreements between the $z$-spin components is due to different choices of the spin length, as $S=\sqrt{1/4+1/(2N)}$ was chosen for the TWA to account for the finite cluster size. These deviations should decrease with increasing $N$, as can be seen in (c). There we show the same comparison for the case of two clusters, $N=60$ and otherwise identical parameters, leading to an even better agreement. We note that $\pm2$ standard errors from the mean is shown by the shaded regions for the HOPS calculations, but is not plotted for the TWA calculations as the error bands are narrower than the lines.}
    \label{fig:TWA_HOPS_3}
\end{figure*}
\subsection{Comparison with the Hierarchy of Pure States}

To check the validity of the results in the previous section, especially for small cluster sizes $N\leq 100$, we compare the TWA against a numerically exact method in the following.
For this task, we use the hierarchy of pure states (HOPS) \cite{suess_hierarchy_2014,hartmann_exact_2017}, which is a quantum trajectory method for non-Markovian Gaussian environments, derived from the non-Markovian state diffusion framework~\cite{diosi_non-markovian_1998}.
As a quantum trajectory method, HOPS is particularly well suited to deal with many-body system ~\cite{flannigan_many-body_2022,Gera2025Jun}. 
Here, we specifically use a matrix product state (MPS) implementation of the recently developed nearly unitary hierarchy of pure states (nuHOPS) method~\cite{muller_quantum_2025}, which is an improved version of HOPS that can successfully deal with highly excited environments such as those encountered in the superradiant phase of the Dicke model~\cite{muller_genuine_2025}. 
The multimode Dicke model can be re-framed as a non-Markovian system, by considering the atomic clusters as the system and the cavity modes as their non-Markovian environment. A detailed discussion of the nuHOPS simulations can be found in Appendix~\ref{app:nuHOPS}.\\

Here, we focus our discussion on the comparison between TWA and nuHOPS and in particular on the example of oscillations in the normal phase. Equivalent comparisons have been made for the superradiant phase as well, showing similar results. In Fig.~\ref{fig:TWA_HOPS_3}(a) and (b) we consider a system made up of three clusters and $N=14$ atoms per cluster. We compare the expectation value of the spin components for the spin clusters calculated using the TWA against the same values calculated using nuHOPS. The values obtained using the two different methods agree remarkably well for $\langle \hat{S}^x_i\rangle/N$ and $\langle \hat{S}^y_i\rangle/N$. The small discrepancy between the initial value of these two quantities can be attributed to the $\mathcal{O}(1/N)$ term in the initial condition Eq.~\eqref{eq:dist_spin} which is not accounted for when sampling the initial spin distribution in the TWA. The larger discrepancy in $\langle\hat{S}^z_i\rangle$ is present due to the use of different spin lengths in the TWA ($S=\sqrt{1/4+1/(2N)}$) and in nuHOPS ($S=1/2$). The difference is particularly noticeable here due to the small number of atoms per cluster. 

We performed a similar comparison on a system made up of only two clusters with $N=60$, in which case the agreement between nuHOPS and TWA is even better, as expected. The good agreement establishes the validity of the results obtained in Sec.~\ref{sec:finite-size-clusters} and thus of the conclusion that persistent oscillations and multistabilty can be observed in finite-sized clusters for finite times. This in turn also helps validate the atom-only approximation made in the derivation of the mean-field equations.

\section{Potential Experimental Implementations}\label{sec:expt_implementations}
One platform for implementing this model experimentally is the multimode cavity QED setup of~\cite{kroeze_directly_2025,marsh_high-capacity_2025,kroeze_high_2023,guo_emergent_2019,vaidya_tunable-range_2018,guo_optical_2021,kollar_adjustable-length_2015,kollar_supermode-density-wave-polariton_2017} 
which has already been used to implement a multimode Dicke model in order to realize replica symmetry breaking~\cite{kroeze_directly_2025} as well as an associative memory~\cite{marsh_high-capacity_2025}. Here we consider to what extent the nearest-neighbor coupling geometry, which is the subject of this paper, could be implemented in such a setup.

In these multimode cavity QED systems, the effective interactions between clusters, $J_{ij}$, are controlled by the positions of the clusters in one half of the mid-plane of the cavity, $\bm{r}_i$. In particular, in a confocal cavity, the effective coupling matrix that is implemented experimentally is
\begin{equation}\label{eq:exp_coupling}
    J_{ij}(\bm{r})\propto g^2\left(\beta\delta_{ij}+\cos\left(2\frac{\bm{r}_i\cdot\bm{r}_j}{w_0^2}\right)\right),
\end{equation}
where $\beta$, the ratio between the strengths of the local interactions and the non-local interactions, is a geometric factor that accounts for the size and shape of the atomic clusters, and $w_0$ is the radius of the Gaussian $\text{TEM}_{00}$ mode~\cite{marsh_quantum-optical_2024,marsh_entanglement_2024,marsh_enhancing_2021}. In the following analysis, we assume a value of $\beta=10$, as suggested by~\cite{marsh_entanglement_2024,marsh_enhancing_2021} to match recent confocal cavity experiments~\cite{vaidya_tunable-range_2018}.

The nearest-neighbor coupling geometry we have considered so far has the following effective coupling matrix:
\begin{equation}\label{eq:nn_coupling}
    J^\mathrm{nn}_{ij}= g^2\begin{cases}
        1, &\text{if } i=j \text{ and } i=1,\ n_c,\\
        2, &\text{if } i=j \text{ and } i\neq1,\ n_c,\\
        1, &\text{if } i=j\pm1,\\
        0, &\text{otherwise},
    \end{cases}
\end{equation} 
which cannot be achieved using the experimental coupling matrix given by Eq.~\eqref{eq:exp_coupling} for $\beta=10$. However, optimizing the cluster positions, $\bm{r}_i$, by minimizing the difference between the nearest-neighbor coupling matrix and the coupling matrix given by Eq.~\eqref{eq:exp_coupling} using the Frobenius norm, $||J^\mathrm{nn}-J(\bm{r})||^2_F=\sum_{i,j=1}^{n_c}|J^\mathrm{nn}_{ij}-J_{ij}(\bm{r})|^2$, results in effective coupling matrices that still lead to multiple stable steady-state solutions in the superradiant phase. For these optimized $J_{ij}$ matrices numerical calculations of the stable steady-state solutions of the mean-field equations suggest that there are
\begin{equation}
    N_\mathrm{sss}(n_c)=2^{n_c},
\end{equation}
\begin{figure}[t!]
  \includegraphics[width=0.9\linewidth]{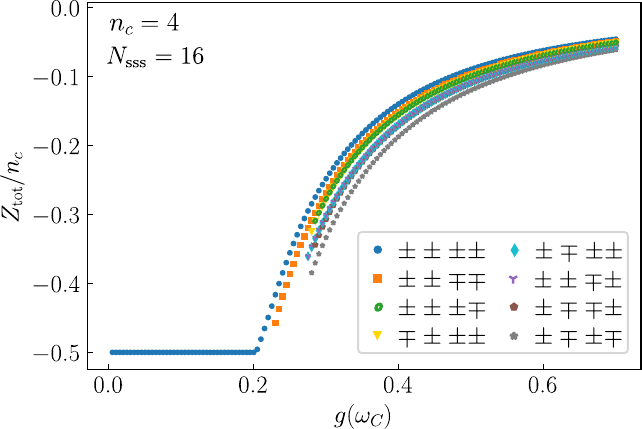}
\caption{Stable steady-state solutions shown through the total $z$-spin components scales by the number of clusters plotted as a function of the coupling strength, $g$, for a $J_{ij}$ effective coupling matrix optimized to match $J^\mathrm{nn}_{ij}$ for four clusters. The legend labels the 8 solution pairs that are present in the superradiant phase. In the optimized $J_{ij}$ matrix the couplings are not fully symmetric around the center so each solution pair corresponds to a distinct set of $Z_\mathrm{tot}$ values. The asymmetry in the couplings is comparatively small so the differences in $Z_\mathrm{tot}$ values can be very small, causing the points plotted in this figure to overlap. The system parameters for these calculations are $\{\omega_A,\kappa\}=\{0.1,1\}\omega_C$ and the optimized $J_{ij}$ matrix and corresponding atomic positions, $\bm{r}_i$, can be found in Appendix~\ref{app:opt_J}.}
  \label{fig:exp_SS}
\end{figure}
stable steady-state solutions in the superradiant phase. As in the nearest-neighbor coupling configuration, this also comes from identifying a composition-based pattern in the pairs of stable steady-state solutions, except, that in this case all compositions of $n_c$ correspond to stable steady-state solutions including those which contain 1. An example of this is shown in Fig.~\ref{fig:exp_SS} for a system with four clusters, where the optimized $J_{ij}$ matrix and corresponding atomic positions, $\bm{r}_i$, can be found in Appendix~\ref{app:opt_J}. Again, supposing this pattern holds for arbitrary numbers of clusters, the number of stable steady-state solutions would grow exponentially with $n_c$. However, none of the stable steady-state solutions for these optimized effective coupling matrices have additional purely imaginary eigenvalues of the Jacobian, and so there are no persistent oscillations.

In order to obtain the true nearest-neighbor coupling geometry needed to observe the persistent oscillations, we can consider an alternative experimental setup involving atomic clusters trapped, using for example optical tweezers, at the intersections of single-mode cavities as in~\cite{leonard_supersolid_2017,baumgartner_stability_2025,leonard_monitoring_2017,morales_coupling_2018}. We propose a configuration of identical single-mode cavities arranged in a zig-zag pattern so that there are two clusters per cavity and the cavities cross at the locations of the clusters. If the clusters are placed symmetrically around the cavity mid-plane within each cavity and in the same locations in all the cavities, then the equal coupling needed to realize the geometry given by Eq.~\eqref{eq:coupling} could be implemented.

\section{Conclusion}\label{sec:Conc}
In this work we considered the multimode Dicke model with an ordered, nearest-neighbor coupling geometry and numerically characterized the mean-field, stable steady-state solutions fully for up to eight clusters, discovering multistability in the superradiant phase and persistent oscillations in both the normal and superradiant phases. We also proved that the number of stable steady-state solutions in the superradiant phase grows as the Fibonacci sequence with the number of clusters in the system. These findings establish that ordered multimode Dicke models can also lead to rich physics, even for a relatively simple coupling geometry, paving the way for further studies of such models. Moreover, we show that these features of the stable-steady state solutions persist for finite-size clusters, where first order quantum fluctuations are taken into account. To do so, we made use of the state-of-the-art numerical method of nuHOPS~\cite{muller_quantum_2025} combined with tensor networks. 

The model is heavily inspired by the multimode cavity QED experiments of~\cite{kroeze_directly_2025,marsh_high-capacity_2025,kroeze_high_2023,guo_emergent_2019,vaidya_tunable-range_2018,guo_optical_2021,kollar_adjustable-length_2015,kollar_supermode-density-wave-polariton_2017} and while the nearest-neighbor coupling geometry is not directly implementable in these setups, it should be feasible to realize it by making use of single-mode cavities intersecting at the positions of atomic clusters~\cite{leonard_supersolid_2017,baumgartner_stability_2025,leonard_monitoring_2017,morales_coupling_2018}. Additionally, in the confocal multimode cavities, a similar, also ordered, effective coupling geometry could be implemented which would also lead to multistability in the superradiant regime. This further emphasizes that ordered coupling geometries, in particular also those without sign-changing couplings, can display interesting features in their steady-state solutions, and that an important direction for future work will be to explore additional ordered coupling geometries to fully understand the effects of the coupling geometry on the steady state solutions of the multimode Dicke model. For example, one could consider imposing periodic boundary conditions on the nearest-neighbor geometry, or changing the number of neighbors by changing the number of modes each cluster couples to. Considering different ordered coupling geometries is particularly interesting in light of the increasing opportunities to couple atoms to multiple cavity modes in varying configurations~\cite{shadmany_cavity_2025,shaw_cavity-array_2026,soper_stability_2026}. With the control now available, carefully arranging clusters within cavities and building bespoke cavity QED systems should allow for the realization of many different coupling geometries in the very near future.

\section*{Acknowledgments}
It is a pleasure to thank J.\ Keeling, P.\ Kirton, A.\ Ziółkowska, C.\ Oliver, and O.\ Adiv for helpful discussions in connection with this work.
The authors gratefully acknowledge the computing time made available to them on the high-performance computer at the NHR Center of TU Dresden. This center is jointly supported by the Federal Ministry of Research, Technology and Space of Germany and the state governments participating in the NHR (www.nhr-verein.de/unsere-partner). The authors would like to acknowledge the use of the University of Oxford Advanced Research Computing (ARC) facility in carrying out this work (https://doi.org/10.5281/zenodo.22558).
M.\ J.\ L.\ and A.\ J.\ D.\ were supported by EPSRC through Grant No.\ EP/T001062/1.
K.\ M.\ gratefully acknowledges the financial support provided by the Graduate Academy of TUD in the form of a travel grant.
Work at Oxford was supported by EPSRC through programme grant QQQS (EP/Y01510X/1), and grant number EP/Z533713/1.

\section*{Data Availability}
The data that support the findings of this work are
openly available~\cite{dataset}.

\appendix
\section{Perturbative Stability Analysis of Superradiant Steady-state Solutions}\label{app:Perturbative_analysis}

Let us perform a perturbative analysis of the mean-field steady-state equations, Eqs.~\eqref{eq:mf_sys}, in the limit $g^{2}/\kappa \gg \omega_{A}$. After using Eq.~\eqref{eq:mf_x} to conclude that $y_{i} = 0$ in the steady-state, the remaining mean-field equations can be written in the form
\begin{equation}
    \omega_{A} x_{i} + 4 g^{2} V_{0} z_{i} (T \bm{x})_{i} = 0,
\end{equation}
where we have defined the tridiagonal matrix
\begin{equation}
    T = \begin{pmatrix} 1 & 1 &  &  & \\ 1 & 2 & 1 &  & \\ & 1 & \ddots & \ddots & \\ & & \ddots & 2 & 1\\ & & & 1 & 1 \end{pmatrix}.
\end{equation}
The spin-conservation law becomes $x_{i}^{2} + z_{i}^{2} = \frac{1}{4}$, and for the superradiant branch connected to the normal phase one takes $z_{i} < 0$, so that $z_{i} = -\sqrt{\frac{1}{4} - x_{i}^{2}}$. Then the non-oscillatory steady-states are the stationary points of the scalar function
\begin{equation}
    \mathcal{F}(\bm{x}) = -\omega_{A} \sum_{i=1}^{n_{c}} \sqrt{\dfrac{1}{4} - x_{i}^{2}} - 2 g^{2} V_{0} \sum_{i=1}^{n_{c}-1} (x_{i} + x_{i+1})^{2},
\end{equation}
in the sense that $\nabla \mathcal{F} = \bm{0}$ gives the steady-state equations.

\subsection{Singletons}\label{sec:singletons}
We have found that it is useful to classify steady-state solutions by the sign pattern of their $x$ spin components. To that end, for a given sign pattern
\begin{equation}
    \bm{\sigma} = (\sigma_{1}, \dots, \sigma_{n_{c}}), \quad \sigma_{i} \in \{\pm 1\},
\end{equation}
we look for a solution with $\sgn(x_{i}) = \sigma_{i}$. For given $\bm{\sigma}$, we define the counting variable
\begin{equation}
    q_{i} = \bm{1}_{\{i>1 \text{ and } \sigma_{i-1} = \sigma_{i}\}} + \bm{1}_{\{i<n_{c} \text{ and } \sigma_{i+1} = \sigma_{i}\}}.
\end{equation}
For a given site $i$ with sign $\sigma_{i}$, $q_{i}$ counts how many nearest-neighbors of site $i$ have the same sign. Clearly we have $q_{i} \in \{0,1,2\}$. It will be useful to define the notion of a \textit{singleton}, meaning a site $i$ with $q_{i} = 0$. For example, the pattern $++-++$ has a singleton $-$ at site 3, whereas $+++--$ has no singletons.

\subsection{Perturbation theory of no-singleton solutions}\label{sec:perturbation_singletons}
In Section.~\ref{sec:stability_anlaysis} of the main text we found that the stable steady-state solutions corresponded to the sign patterns with no singletons. The number of such patterns is given by the Fibonacci numbers, as per Eq.~\eqref{eq:fibonacci}. Let us now argue for the perturbative continuity of these solutions in the strong coupling limit. For a given sign pattern $\bm{\sigma}$, we set
\begin{equation}
    x_{i}(z_{i} ; \sigma_{i}) = \sigma_{i} \sqrt{\dfrac{1}{4} - z_{i}^{2}}, \qquad y_{i} = 0, \qquad z_{i} < 0,
\end{equation}
and then rewrite the steady-state equations in the form
\begin{equation}
    4 z_{i} \big(T \bm{x}(\bm{z};\bm{\sigma})\big)_{i} + \epsilon \, \sigma_{i} \sqrt{\dfrac{1}{4} - z_{i}^{2}} = 0,
    \label{eq:steady_state_z}
\end{equation}
where we have defined the dimensionless parameter
\begin{equation}
    \epsilon = \dfrac{\omega_{A}}{g^{2} V_{0}}.
\end{equation}

We will perform a perturbative expansion in $\epsilon$. First we claim that at $\epsilon = 0$ the no-singleton fixed points all have $\bm{z} = \bm{0}$. Indeed, it is easy to check that
\begin{equation}
    \big(T\bm{x}(\bm{0} ; \bm{\sigma})\big)_{i} = \sigma_{i} q_{i},
\end{equation}
so then the steady-state equations at $\epsilon = 0$ become $4 z_{i} \sigma_{i} q_{i} = 0$. No-singleton solutions have $q_{i} > 0$ for all $i$, so the only solution is $\bm{z} = \bm{0}$, and then $\bm{x} = \bm{\sigma}/2$.

Now we evaluate the Jacobian matrix $\tilde{J}$ of the steady-state equations at $\epsilon = 0$ and $\bm{z} = \bm{0}$. Since $x_{i}(z_{i};\sigma_{i})$ depends on $z_{i}$ only quadratically, we have $\partial x_{i} / \partial z_{j} = 0$ at $\bm{z} = \bm{0}$ for all $i,j$. Therefore, the Jacobian matrix is diagonal with entries
\begin{equation}
    \tilde{J}_{ii} = 4 \big(T\bm{x}(\bm{0} ; \bm{\sigma})\big)_{i} = 4 \sigma_{i} q_{i}.
\end{equation}
Since $q_{i} > 0$ for all $i$ in a non-singleton solution, we have $\tilde{J}_{ii} \neq 0$ for all $i$, so $\tilde{J}$ is an invertible matrix. Therefore, by the analytic implicit function theorem, for every no-singleton sign pattern $\bm{\sigma}$, there exists $\epsilon_{0}(\bm{\sigma}) > 0$ such that for all $\epsilon < \epsilon_{0}(\bm{\sigma})$ there is a unique solution $\bm{z}(\epsilon ; \bm{\sigma})$ to the steady-state equations that is analytic in $\epsilon$, thereby justifying the perturbative expansion. Expanding \cref{eq:steady_state_z} to second order in $\epsilon$ gives
\begin{equation}\label{eq:perturbative_z}
    z_{i}(\epsilon) = -\dfrac{1}{8 q_{i}} \epsilon + \mathcal{O}(\epsilon^{3}),
\end{equation}
and from $x_{i} = \sigma_{i} \sqrt{\frac{1}{4} - z_{i}^{2}}$ we get
\begin{equation}
    x_{i}(\epsilon) = \dfrac{\sigma_{i}}{2} - \dfrac{\sigma_{i}}{64 q_{i}^{2}} \epsilon^{2} + \mathcal{O}(\epsilon^{4}).
\end{equation}

In Sec.~\ref{sec:existence} of this Appendix, we provide an alternative proof of the existence of the no-singleton steady-state solutions, using a fixed point argument which does not rely on a perturbative expansion. This alternative proof allows us to obtain an upper bound on $\epsilon$ for which all the no-singleton steady-states exist.

\subsection{Linearized stability of no-singleton solutions}\label{sec:perturbation_stability} 
To evaluate the stability of the no-singleton solutions, we must linearize the full mean-field dynamics. We begin by recording the fixed-point Hessian of $\mathcal{F}(\bm{x})$, which will determine the stiffness matrix in the linearized dynamics. We have
\begin{equation}
    H_{\mathcal{F}}(\bm{x}) = \mathrm{diag}\left(\dfrac{-\omega_{A}}{4 z_{i}^{3}}\right) - 4 g^{2} V_{0} T,
\end{equation}
where we have computed $\partial^{2} \mathcal{F} / \partial x_{i} \partial x_{j}$ and used the steady-state equations to rewrite the first term in terms of $z_{i}$. Using the perturbative expansion for $z_{i}(\epsilon)$, given by Eq.~\eqref{eq:perturbative_z}, we can write this as 
\begin{equation}
    H_{\mathcal{F}}(\bm{x}) = \mathrm{diag}\left(128 \omega_{A} q_{i}^{3} \epsilon^{-3} + \mathcal{O}(\epsilon^{-1}) \right) - 4 \omega_{A} \epsilon^{-1} T,
\end{equation}
Standard results on tridiagonal matrices~\cite{yueh_eigenvalues_2005}, show that the eigenvalues of $T$ are given by $\lambda_{m} = 4 \cos^{2}\left(m \pi / 2 n_{c}\right)$ for $m = 1, \dots, n_{c}$, so $\norm{T} \leq 4$, independent of $n_{c}$. Then, since no-singleton solutions have $q_{i} > 0$ for all $i$, we have that the Hessian is positive definite for sufficiently small $\epsilon$, so the no-singleton solutions are local minima of $\mathcal{F}$.

Now let $(\bm{x}_{*}, \bm{0}, \bm{z}_{*})$ be one of the no-singleton steady-states constructed above, and write perturbations as $(\delta \bm{x}, \delta \bm{y}, \delta \bm{z})$. Linearizing the full mean-field equations, Eqs.~\eqref{eq:mf_sys}, around this fixed point gives
\begin{align}
    \delta \dot{\bm{x}} =& -\omega_{A} \delta \bm{y}, \\
    \delta \dot{y}_{i} =&\, \omega_{A} \delta x_{i}  + 4 g^{2} V_{0} z_{*,i} (T \delta \bm{x})_{i} + 4 g^{2} V_{0} (T \bm{x}_{*})_{i} \delta z_{i} \notag \\ &+ 4 g^{2} V_{1} z_{*,i} (T \delta \bm{y})_{i}, \\
    \delta \dot{z}_{i} =& -4 g^{2} V_{0} (T\bm{x}_{*})_{i} \delta y_{i}.
\end{align}

Because the spin length is conserved, tangent perturbations satisfy $x_{*,i} \delta x_{i} + z_{*,i} \delta z_{i} = 0$, and hence
\begin{equation}
    \delta z_{i} = - \dfrac{x_{*,i}}{z_{*,i}} \delta x_{i}.
\end{equation}
Using the steady-state equation for $(\bm{x}_{*}, \bm{z}_{*})$ together with $x_{*,i}^{2} + z_{*,i}^{2} = \frac{1}{4}$, a few lines of algebra show that this reduces the linearized system to
\begin{equation}
    \delta \dot{\bm{y}} = P H_{\mathcal{F}}(\bm{x}_{*}) \delta \bm{x} - 4 g^{2} V_{1} P T \delta \bm{y},
\end{equation}
where
\begin{equation}
    P = \mathrm{diag}(-z_{*,1}, \dots, -z_{*,n_{c}}).
\end{equation}
Since $z_{*,i} < 0$, the matrix $P$ is positive definite. Eliminating $\delta \bm{y} = - \omega_{A}^{-1} \delta \dot{\bm{x}}$ gives the second-order system
\begin{equation}
    \delta \ddot{\bm{x}} + 4 g^{2} V_{1} P T \delta \dot{\bm{x}} + \omega_{A} P H_{\mathcal{F}}(\bm{x}_{*}) \delta \bm{x} = 0.
\end{equation}
If we now write $\delta \bm{x} = P^{1/2} \bm{u}$, then $\bm{u}$ satisfies the equation of motion for a generalized damped harmonic oscillator,
\begin{equation}
    \ddot{\bm{u}} + 4 g^{2} V_{1} G \dot{\bm{u}} + \omega_{A} K \bm{u} = 0,
    \label{eq:u_ode}
\end{equation}
where $G$, the damping matrix, and $K$, the stiffness matrix, are given by
\begin{equation}
    G = P^{1/2} T P^{1/2}, \qquad K = P^{1/2} H_{\mathcal{F}}(\bm{x}_{*}) P^{1/2}.
\end{equation}
Since $P$ is a diagonal, positive definite ($P\succ0$) matrix, we can always choose the square root $P^{1/2}$ to also be diagonal and positive definite (and hence symmetric). Then $T \succeq 0$ implies $G \succeq 0$, and since $H_{\mathcal{F}}(\bm{x}_{*}) \succ 0$ for small enough $\epsilon$, we have $K \succ 0$ in the same regime.

To determine the Lyapunov stability of the system, we consider the quadratic energy function
\begin{equation}
    E(\bm{u}, \dot{\bm{u}}) = \dfrac{1}{2} \norm{\dot{\bm{u}}}^{2} + \dfrac{\omega_{A}}{2} \bm{u}^{T} K \bm{u}.
\end{equation}
In the regime of $\epsilon$ sufficiently small that the stiffness matrix $K$ is positive definite, we have $E \geq 0$, with equality if and only if $\bm{u} = \dot{\bm{u}} = \bm{0}$. Furthermore, \cref{eq:u_ode} implies that $\dot{E}$ satisfies
\begin{equation}
    \dot{E} = - 4 g^{2} V_{1} \dot{\bm{u}}^{T} G \dot{\bm{u}}.
\end{equation}
A direct computation from the definitions of $\alpha_{\pm}$, given in Eq.~\eqref{eq:alpha_pm}, gives
\begin{equation}
    V_{1} = \dfrac{4 \kappa \, \omega_{C} \, \omega_{A}}{\big((\omega_{C}+\omega_{A})^{2}+\kappa^{2}\big)\big((\omega_{C}-\omega_{A})^{2}+\kappa^{2}\big)} > 0.
    \label{eq:V1_pos}
\end{equation}
Since $G \succeq 0$, we therefore have $\dot{E} \leq 0$, so $E$ is a Lyapunov function, and we conclude that the fixed point is Lyapunov stable for $\epsilon$ sufficiently small~\cite{khalil_nonlinear_2002}.

\subsection{Persistent oscillations in the Superradiant phase}\label{sec:perturbation_oscillations}
To determine when the stability is asymptotic (meaning the system converges to the equilibrium point as $t \to \infty$), we use LaSalle's invariance principle~\cite{khalil_nonlinear_2002}. The kernel of $T$ is one-dimensional and spanned by the alternating vector $\bm{a} = (1,-1,1,-1,\dots)^{T}$, so the only potentially undamped direction is $\ker G = \mathrm{span}\{P^{-1/2}\bm{a}\}$. A neutral oscillatory mode can only persist if this direction is also an eigenvector of the stiffness matrix $K$. Since $T \bm{a} = \bm{0}$, this requires
\begin{equation}
    P H_{\mathcal{F}}(\bm{x}_{*}) \bm{a} = \mathrm{diag}\left(\dfrac{\omega_{A}}{4 z_{*,i}^{2}}\right) \bm{a} \overset{!}{=} \lambda \bm{a},
\end{equation}
for some eigenvalue $\lambda$, which happens if and only if $|z_{*,i}|$ is independent of $i$. On the perturbative branch we have
\begin{equation}
    z_{*,i} = -\dfrac{1}{8 q_{i}} \epsilon + \mathcal{O}(\epsilon^{3}),
\end{equation}
so this requires $q_{i}$ to be independent of $i$. For a no-singleton pattern this means $q_{i} = 1$ for all $i$, i.e. the sign pattern consists of alternating blocks of length two. These are precisely the cases in which Section~\ref{sec:persistent_oscillations_SR} of the main text finds persistent oscillations. We cannot have all $q_{i} = 2$ because the boundary clusters only have one neighbor, and hence have $q_{i} \leq 1$. For every other no-singleton sign pattern, the invariant set is therefore empty and the linearized steady-state is asymptotically stable.

\subsection{Scaling of the perturbation parameter} Let us now examine the physical conditions under which we have $\epsilon = \omega_{A} / g^{2} V_{0} \ll 1$. An explicit computation shows that $V_{0} = \Re(\alpha_{+})$ is given by
\begin{equation}
    V_{0} = \dfrac{2 \omega_{C} (\omega_{c}^{2} + \kappa^{2} - \omega_{A}^{2})}{(\omega_{C}^{2} + \kappa^{2} - \omega_{A}^{2})^{2} + 4 \omega_{A}^{2} \kappa^{2}}.
\end{equation} 
Assuming $\omega_{C} \sim \kappa$, and writing $g = r_{g} \kappa$ and $\omega_{A} = r_{A} \kappa$, we then have
\begin{equation}
    \epsilon = \mathcal{O}\left(\dfrac{r_{A}}{r_{g}^{2}}\right),
\end{equation}
so for $\epsilon \ll 1$ we need $r_{g}^{2} \gg r_{A}$, which is equivalent to the condition
\begin{equation}
    \dfrac{g^{2}}{\kappa} \gg \omega_{A}.
\end{equation}
We emphasize that this is compatible with both $g \ll \kappa, \omega_{C}$ and $\omega_{A} \ll \kappa, \omega_{C}$, which are conditions for the validity of the atom-only description under which we derived the mean-field equations.

\subsection{Existence of all no-singleton stable steady-states for small $\epsilon$}\label{sec:existence}
Finally, we find an upper bound on the value of $\epsilon$ needed to guarantee that the Hessian, $H_{\mathcal{F}}(\bm{x}_{*})$, is positive definite, and thus to guarantee that all the no-singleton stable steady-state solutions exist.

The exact Hessian is given by
\begin{equation}
    H_{\mathcal{F}}(\bm{x}_{*}) = \omega_{A}\left[\mathrm{diag}\left(\dfrac{1}{4|z_{*,i}|^{3}}\right) - \dfrac{4}{\epsilon} T\right].
\end{equation}
To ensure $H_{\mathcal{F}}(\bm{x}_{*})$ is positive definite, it suffices to ensure
\begin{equation}
    \dfrac{1}{4 \norm{\bm{z}_{*}(\epsilon)}_{\infty}^{3}} - \dfrac{16}{\epsilon} > 0,
    \label{eq:positivity_condition}
\end{equation}
where $\norm{\bm{z}_{*}(\epsilon)}_{\infty} = \max_{i} |z_{*,i}(\epsilon)|$, and we used $\lambda_{\mathrm{max}}(T) \leq 4$. So we need to derive an upper bound on $\norm{\bm{z}_{*}(\epsilon)}_{\infty}$ from the mean-field equations. Below we will prove that we can get a suitable upper bound provided we take
\begin{equation}
    \boxed{\epsilon \leq \dfrac{3}{4}.}
\end{equation}
Note that, while this is explicit, it is unlikely to be tight since there are various numerical factors in the proof that we have not attempted to optimize.

\paragraph*{Steady state solution as a fixed point.}
Recall that at $\bm{z} = 0$ we have $\sigma_{i} (T \bm{x}(\bm{0};\bm{\sigma}))_{i} = q_{i}$. That motivates the definition
\begin{equation}
    \delta_{i}(\bm{z};\bm{\sigma}) = \sigma_{i} (T \bm{x}(\bm{z};\bm{\sigma}))_{i} - q_{i} = \sigma_{i} \big(T[\bm{x}(\bm{z};\bm{\sigma}) - \bm{x}(\bm{0};\bm{\sigma})]\big)_{i},
\end{equation}
in terms of which we can write the mean-field steady-state equations, Eq.~\eqref{eq:steady_state_z}, as
\begin{equation}
    z_{i} = -\dfrac{\delta_{i}}{q_{i}} z_{i} - \dfrac{\epsilon}{4 q_{i}} \sqrt{\frac{1}{4} - z_{i}^{2}},
\end{equation}
where we have assumed a no-singleton solution so that we can divide by $q_{i} \neq 0$. Then we define the map $\Psi_{\epsilon}$ by
\begin{equation}
    \big(\Psi_{\epsilon}(\bm{z})\big)_{i} = -\dfrac{\delta_{i}}{q_{i}} z_{i} - \dfrac{\epsilon}{4 q_{i}} \sqrt{\frac{1}{4} - z_{i}^{2}},
\end{equation}
so that the steady-state solution $\bm{z}_{*}(\epsilon)$ is a fixed point of $\Psi_{\epsilon}$. The goal will be to apply the Banach fixed point theorem to $\Psi_{\epsilon}$ on a suitable space to be defined shortly. This will provide an alternative route to proving the existence of the steady-state solution as compared to appealing to the invertibility of the Jacobian, with the advantage that we can be more explicit about bounding $\epsilon$.

Now define the ball
\begin{equation}
    B_{r} = \{\bm{z} \in \mathbb{R}^{n_{c}} : \norm{\bm{z}}_{\infty} \leq r\}.
\end{equation}
Since $\bm{x}(\bm{z}; \bm{\sigma})_{i} = \sigma_{i} \sqrt{\frac{1}{4} - z_{i}^{2}}$, we have
\begin{equation}
    \delta_{i}(\bm{z}; \bm{\sigma}) = \sum_{j} \sigma_{i} T_{ij} \sigma_{j} \left(\sqrt{\dfrac{1}{4} - z_{j}^{2}} - \dfrac{1}{2}\right).
\end{equation}
For $\bm{z} \in B_{r}$, we have the bound
\begin{equation}
    \dfrac{1}{2} - \sqrt{\dfrac{1}{4} - z_{j}^{2}} = \dfrac{z_{j}^{2}}{\frac{1}{2} + \sqrt{\frac{1}{4} - z_{j}^{2}}} \leq 2r^{2},
\end{equation}
which implies
\begin{equation}
    |\delta_{i}(\bm{z}; \bm{\sigma})| \leq 2 r^{2} \sum_{j} |T_{ij}| \leq 8 r^{2},
\end{equation}
where we used $\sum_{j} |T_{ij}| \leq 4$. Applying this to $\Psi_{\epsilon}$, we have
\begin{equation}
    |\big(\Psi_{\epsilon}(\bm{z})\big)_{i}| \leq 8 r^{3} + \dfrac{\epsilon}{8}.
\end{equation}
If we set $r = \epsilon / 4$ for any $0 < \epsilon < 1$, then
\begin{equation}
    |\big(\Psi_{\epsilon}(\bm{z})\big)_{i}| \leq \dfrac{1}{8}(\epsilon^{3} + \epsilon) \leq \dfrac{\epsilon}{4} = r,
\end{equation}
so $\Psi_{\epsilon}$ maps $B_{\epsilon/4}$ to itself.

\paragraph*{Contraction.} 
Next we need to establish that $\Psi_{\epsilon}$ is a contraction on $B_{\epsilon/4}$ for small enough $\epsilon$, meaning that there exists $\kappa < 1$ such that $\norm{\Psi_{\epsilon}(\bm{z}) - \Psi_{\epsilon}(\bm{w})}_{\infty} \leq \kappa \norm{\bm{z} - \bm{w}}_{\infty}$ for all $\bm{z}, \bm{w} \in B_{\epsilon/4}$. We have
\begin{multline}
        \big(\Psi_{\epsilon}(\bm{z}) - \Psi_{\epsilon}(\bm{w})\big)_{i} = -\dfrac{1}{q_{i}}\big(z_{i} \delta_{i}(\bm{z};\bm{\sigma}) - w_{i} \delta_{i}(\bm{w};\bm{\sigma})\big) \\- \dfrac{\epsilon}{4 q_{i}} \left(\sqrt{\dfrac{1}{4} - z_{i}^{2}} - \sqrt{\dfrac{1}{4} - w_{i}^{2}}\right).
    \label{eq:Psi_diff}
\end{multline}

To bound the RHS, we first notice that
\begin{align}
    \left|\sqrt{\dfrac{1}{4} - z_{i}^{2}} - \sqrt{\dfrac{1}{4} - w_{i}^{2}}\right| &= \dfrac{|z_{i}^{2} - w_{i}^{2}|}{\sqrt{\frac{1}{4} - z_{i}^{2}} + \sqrt{\frac{1}{4} - w_{i}^{2}}}, \\ &= \dfrac{|z_{i}+w_{i}||z_{i}-w_{i}|}{\sqrt{\frac{1}{4} - z_{i}^{2}} + \sqrt{\frac{1}{4} - w_{i}^{2}}}.
\end{align}
Since $\bm{z}, \bm{w} \in B_{\epsilon/4}$, we have $|z_{i}+w_{i}| \leq \epsilon/2$, and provided $\epsilon < 1$ we have $\sqrt{\frac{1}{4} - z_{i}^{2}} + \sqrt{\frac{1}{4} - w_{i}^{2}} \geq 2 \sqrt{\frac{1}{4} - \frac{\epsilon^{2}}{16}} \geq 1/2$, which together imply
\begin{equation}
    \left|\sqrt{\dfrac{1}{4} - z_{i}^{2}} - \sqrt{\dfrac{1}{4} - w_{i}^{2}}\right| \leq \epsilon \norm{\bm{z} - \bm{w}}_{\infty}.
    \label{eq:sqrt_diff}
\end{equation}
This provides a bound for the second factor on the RHS of \cref{eq:Psi_diff}. To bound the first factor, we write $z_{i} \delta_{i}(\bm{z};\bm{\sigma}) - w_{i} \delta_{i}(\bm{w};\bm{\sigma}) = z_{i} \big(\delta_{i}(\bm{z};\bm{\sigma}) - \delta_{i}(\bm{w};\bm{\sigma})\big) + (z_{i} - w_{i}) \delta_{i}(\bm{w};\bm{\sigma})$. We already have the bound $|\delta_{i}(\bm{w};\bm{\sigma})| \leq 8 r^{2}$ for the second factor, while
\begin{align}
    |\delta_{i}(\bm{z};\bm{\sigma}) -& \delta_{i}(\bm{w};\bm{\sigma})| \\&= \left|\sum_{j} \sigma_{i} T_{ij} \sigma_{j} \left(\sqrt{\dfrac{1}{4} - z_{j}^{2}} - \sqrt{\dfrac{1}{4} - w_{j}^{2}}\right)\right|\nonumber \\& \leq 4\epsilon \norm{\bm{z} - \bm{w}}_{\infty},
\end{align}
where we used \cref{eq:sqrt_diff} and $\sum_{j} |T_{ij}| \leq 4$. We can combine these results to get
\begin{equation}
    \left|z_{i} \delta_{i}(\bm{z};\bm{\sigma}) - w_{i} \delta_{i}(\bm{w};\bm{\sigma}) \right| \leq \left(4r \epsilon + 8r^{2}\right) \norm{\bm{z} - \bm{w}}_{\infty}.
\end{equation}
Applying this bound and \cref{eq:sqrt_diff} to \cref{eq:Psi_diff}, we conclude that
\begin{equation}
    \norm{\Psi_{\epsilon}(\bm{z}) - \Psi_{\epsilon}(\bm{w})}_{\infty} \leq \left(4 r \epsilon + 8r^{2} + \dfrac{\epsilon^{2}}{4}\right) \norm{\bm{z} - \bm{w}}_{\infty}.
\end{equation}
With $r = \epsilon/4$, we have
\begin{equation}
    \kappa \equiv 4 r \epsilon + 8r^{2} + \dfrac{\epsilon^{2}}{4} = \dfrac{7}{4} \epsilon^{2}.
\end{equation}
To guarantee $\kappa < 1$ it suffices to take $\epsilon \leq \frac{3}{4}$, since then $\frac{7}{4} \epsilon^{2} \leq \frac{63}{64} < 1$. Hence for $\epsilon \leq \frac{3}{4}$, $\Psi_{\epsilon}$ is a contraction mapping on $B_{\epsilon/4}$, and therefore by Banach's fixed point theorem the solution $\bm{z}_{*}(\epsilon)$ exists as a fixed point of $\Psi_{\epsilon}$ in $B_{\epsilon/4}$.

Having established the existence of the steady-state solution $\bm{z}_{*}(\epsilon)$ provided $\epsilon \leq \frac{3}{4}$, we can finally go back to the positivity criterion \cref{eq:positivity_condition}. Since $\norm{\bm{z}_{*}(\epsilon)}_{\infty} \leq \epsilon/4$, we have
\begin{equation}
    \dfrac{1}{4 \norm{\bm{z}_{*}(\epsilon)}_{\infty}^{3}} - \dfrac{16}{\epsilon} \geq 16(\epsilon^{-3} - \epsilon^{-1}),
\end{equation}
which is indeed greater than zero for any $0 < \epsilon < 1$, so we conclude that the Hessian $H_{\mathcal{F}}(\bm{x}_{*})$ is positive definite, and the remaining stability analysis in Sec.~\ref{sec:perturbation_stability} ensures that this solution is stable.

\section{Derivation of Expressions for Oscillatory Solutions}\label{app:Osc_sols}
To derive the expressions for the oscillatory solutions, both in the superradiant phase (Eqs.~\eqref{eq:osc_sols_SR}) and in the normal phase (Eqs.~\eqref{eq:osc_sols_NP}), we start from the mean-field equations, Eqs.~\eqref{eq:mf_sys}, for two clusters and, as in~\cite{iemini_dynamics_2024}, derive equations of motion for the linear combinations $\beta_\mathrm{tot}=\beta_1+\beta_2$ and $\beta_\mathrm{d}=\beta_1-\beta_2$, with $\beta=x,\ y,\ z$:
\begin{subequations}\label{eq:mf_tot}
\begin{align}
    \dot{x}_\mathrm{tot}&=-\omega_Ay_\mathrm{tot}, \\
    \dot{y}_\mathrm{tot}&=\omega_Ax_\mathrm{tot}+4g^2V_0z_\mathrm{tot}x_\mathrm{tot}+4g^2V_1z_\mathrm{tot}y_\mathrm{tot},\\
    \dot{z}_\mathrm{tot}&=-4g^2V_0y_\mathrm{tot}x_\mathrm{tot}-4g^2V_1y_\mathrm{tot}^2,
\end{align}
\end{subequations}
\begin{subequations}
\begin{align}
    \dot{x}_\mathrm{d}&=-\omega_Ay_\mathrm{d},\\
    \dot{y}_\mathrm{d}&=\omega_Ax_\mathrm{d}+4g^2V_0z_\mathrm{d}x_\mathrm{tot}+4g^2V_1z_\mathrm{d}y_\mathrm{tot},\\
    \dot{z}_\mathrm{d}&=-4g^2V_0y_\mathrm{d}x_\mathrm{tot}-4g^2V_1y_\mathrm{d}y_\mathrm{tot}.
\end{align}
\end{subequations}
We can see that Eqs.~\eqref{eq:mf_tot} for the $\beta_\mathrm{tot}$ variables are self consistent, and can be solved for steady-state solutions. These closely resemble the mean-field equations of the standard Dicke model, and we find that:
\begin{align}
    y_\mathrm{tot}^\mathrm{ss}=0, && x_\mathrm{tot}^\mathrm{ss}=0, && z_\mathrm{tot}^\mathrm{ss}=-S,
\end{align}
for the normal phase, or 
\begin{align}
        y_\mathrm{tot}^\mathrm{ss}&=0, \nonumber \\ x_\mathrm{tot}^\mathrm{ss}&=\pm\sqrt{S^2-\omega_A^2/16g^4V_0^2},  \\  z_\mathrm{tot}^\mathrm{ss}&=-\omega_A/4g^2V_0 , \nonumber
\end{align}
for the superradiant phase, where $S^2=x_\mathrm{tot}^2+y_\mathrm{tot}^2+z_\mathrm{tot}^2$ is fixed by the initial conditions as it is a conserved quantity. Substituting these into the equations for the $\beta_\mathrm{d}$ variables, allows their long-time (oscillatory) behavior to be determined. From there we can find the desired expressions for the oscillations carried out by the individual spin clusters.

\section{Details of the nuHOPS Simulations}\label{app:nuHOPS}
As described in the main text, the multimode Dicke model can be understood as a system in contact with a non-Markovian bath by considering the spins as the system of interest $\hat{\rho}_\mathrm{sys} = \Tr[\mathrm{cavity}]{\hat{\rho}}$ and the cavity modes as the (in general non-Markovian) environment. 
Taking a Markovian limit of this environment \cite{jager_lindblad_2022} leads to the atom-only environment discussed in Sec.~\ref{ssec:atom_only}.
Here, we proceed without this approximation.
By defining the system Hamiltonian $\hat{H}_\mathrm{sys} = \omega_A\sum_{i=1}^{N_{s}}\hat{S}_i^z$ and the coupling operators $\hat{L}_j^\mathrm{tot} = \sum_{i=1}^{N_s} g_{ij}\hat{S}_i^x$ it is straightforward to set up the corresponding nuHOPS equation from \cite{muller_quantum_2025} as 
\begin{equation}\label{eq:nuhopsMultiDicke}
    \begin{split}
        \partial_t |\Phi(t)\rangle =& -i\hat{H}_\mathrm{sys}\\
        &+ \sum_{j=1}^{n_m}\Bigg[(\nu_j^*(t)\hat{L}_j^\mathrm{tot}-\nu_j(t)\hat{L}_j^{\mathrm{tot}\,\dagger}) + -i\omega_c\hat{b}_j^\dagger \hat{b}_j\\
        &\quad\quad\quad\quad-i\left(\hat{L}_j^\mathrm{tot} - \langle \hat{L}_j^\mathrm{tot}\rangle_t\right)\hat{b}_j^\dagger \\
        &\quad\quad\quad\quad-i  \left(\hat{L}_j^{\mathrm{tot}\,\dagger} - \langle \hat{L}_j^{\mathrm{tot}\,\dagger}\rangle_t\right)\hat{b}_j \\
        &\quad\quad\quad\quad+ {\eta}_{j}^*(t)\hat{L}_j^\mathrm{tot} -\kappa \hat{b}_j^\dagger \hat{b}_j \Bigg] |\Phi(t)\rangle,\\
        \dot{\nu}_j =& -(i\omega_c + \kappa)\nu_j + \langle \hat{L}_j^\mathrm{tot}\rangle_t,
    \end{split}
\end{equation}
where $\hat{b}_j$ ($\hat{b}_j^\dagger$) are bosonic annihilation (creation) operators and the noises ${\eta}_{j}^*(t)$ are independent, identically distributed Gaussian random processes with $\mean{{\eta}_{j}^*(t)} = 0$ and $\mean{{\eta}_{i}(t){\eta}_{j}^*(s)} = \delta_{ij}\exp{-i\omega_c(t-s)-\kappa|t-s|}$.
It is important to note that the trajectories of the physical (spin-) system are obtained from a projection onto the oscillator vacuum state $|\psi_t(\eta^*)\rangle = \langle0|\Phi(t)\rangle$.
The desired reduced state is then obtained from the trajectory average
\begin{equation}
    \hat{\rho}_\mathrm{sys} = \mean{\frac{|\psi_t(\eta^*)\rangle\!\langle\psi_t(\eta^*)|}{\langle\psi_t(\eta^*)|\psi_t(\eta^*)\rangle}}.
\end{equation}
Furthermore, all expectation values in Eq.~\eqref{eq:nuhopsMultiDicke} are taken with respect to this physical trajectory
\begin{equation}
    \langle \hat{L}_j^\mathrm{tot}\rangle_t = \frac{\langle\psi_t(\eta^*)|\hat{L}_j^\mathrm{tot}|\psi_t(\eta^*)\rangle}{\langle \psi_t(\eta^*)|\psi_t(\eta^*)\rangle}.
\end{equation}
As explained in Ref.~\cite{muller_quantum_2025}, Eq.~\eqref{eq:nuhopsMultiDicke} is constructed in a way that achieves $\langle\Phi(t)|\hat{b}_j|\Phi(t)\rangle/\langle\Phi(t)|\Phi(t)\rangle \approx 0$ throughout the dynamics, which allows for an efficient Fock state expansion of the auxiliary oscillators.
An intuitive picture for the physical meaning of the state $|\Phi(t)\rangle$ can be obtained from a comparison to the pseudomode method \cite{Muller2026Apr}.\\
\begin{figure}[t!]
    \centering
    \includegraphics[width=0.8\linewidth]{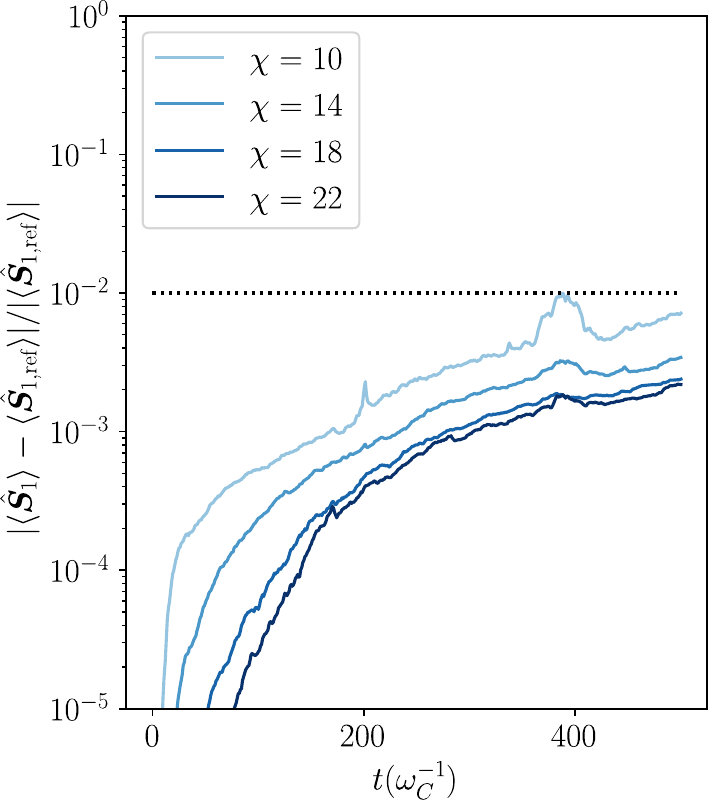}
    \caption{Convergence checks for the HOPS simulations for three clusters shown in Fig.~\ref{fig:TWA_HOPS_3} (a) and (b). To asses the error we compared the deviations in the spin expectation values in the first cluster $\langle \bm{S}_1\rangle$ for different values of $\chi$ and $k_\mathrm{max}=8$ to a reference calculation $\langle \bm{S}_{1,ref}\rangle$, which was obtained with $\chi=26,\,k_{max}=18$. The plot shows the single trajectory errors averaged over 100 trajectories for different bond dimensions $\chi$.}
    \label{fig:hops_convergence}
\end{figure}

To solve Eq.~\eqref{eq:nuhopsMultiDicke} we make an MPS Ansatz for $|\Phi(t)\rangle$ as 
\begin{align}
    |\Phi&(t)\rangle = \label{eq:mpsAnsatzPhi}\\
    &\sum_{\bm{m},\bm{n}}  C^{m_1}[1]C^{n_1}[2] \dots C^{m_{N_s}}[2N_s-1]|m_1,n_1,\dots, m_{N_s}\rangle,\notag
\end{align}
where $\{|m_i\rangle\}$ are the basis states of spin $i$, $\{|n_j\rangle\}$ are Fock states of the auxiliary oscillator $j$ and $C^{a}[b]$ are $\chi\times\chi$ matrices.
To propagate the Ansatz we use a variant of the time-dependent variational principle (TDVP) \cite{Haegeman2011Aug,Haegeman2016Oct}.
For this it is important to consider, that, in contrast to the usual Schrödinger equation, nuHOPS is a non-linear differential equation.
This non-linearity cannot be combined with Trotter decomposition in the most commonly used one-site and two-site TDVP algorithms \cite{Haegeman2016Oct}. Instead, we need to resort to the TDVP algorithm for a fixed MPS representation \cite{Haegeman2011Aug, Li2020Jan}.
The differential equations for the stochastic process shifts $\nu_j(t)$ can simply be added to the set of differential equations for the MPS parameters and propagated alongside them.
For the implementation we make use of the python based \texttt{renormalizer} library \cite{renormalizer,ren2018time, Li2020Jan}.
As discussed in Ref.~\cite{Li2020Jan}, this library uses a regularization \cite{Meyer2018Mar} to prevent the instabilities that can occur in the TDVP algorithm for a fixed MPS representation.
Our simulations are based the \texttt{tdvp\_vmf} method implemented in \texttt{renormalizer}, where we have made some adjustments to propagate the stochastic process shifts alongside the wavefunction.\\

For the simulations shown in Fig.~\ref{fig:TWA_HOPS_3} we used a bond dimension $\chi=18$ and a Fock-state truncation $k_{max}=8$ for the three cluster case (plots (a) and (b)) and a bond dimension $\chi=13$, $k_{max}=22$ for the two cluster case (plot(c)). The parameters were chosen following a convergence check as shown in Fig.~\ref{fig:hops_convergence} for the three cluster case.

\section{Optimized Effective Coupling Matrix for 4 Clusters}\label{app:opt_J}
\begin{figure}[h!]
  \includegraphics[width=0.9\linewidth]{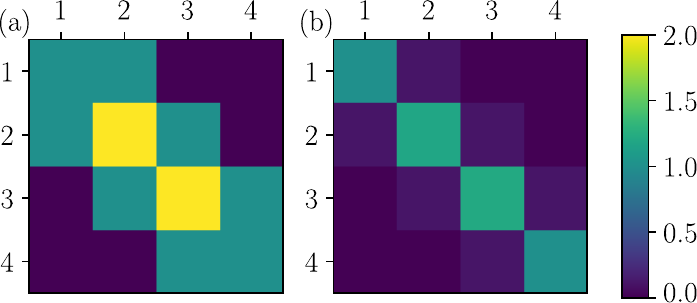}
\caption{Coupling matrices for (a) the nearest-neighbor coupling geometry given by Eq.~\eqref{eq:nn_coupling} and (b) the optimized coupling matrix in Eq.~\eqref{eq:coupling_opt}. The matrix in (b) has been scaled by the value of $J_{11}$ for a clearer comparison. }
  \label{fig:Jij_comparison}
\end{figure}

In Fig.~\ref{fig:exp_SS} we show the stable steady-state solutions obtained using the experimentally viable effective coupling matrix, $J_{ij}$ given by Eq.~\eqref{eq:exp_coupling}, for which the atomic positions $\bm{r}_i$ were optimized to match $J^\mathrm{nn}_{ij}$ for four clusters. The optimized atomic positions (rounded to 6dp), which are chosen to lie in the same half of the mid-plane of the cavity, are
\begin{align}
\begin{split}
    (r_1^x,r_1^y)&=(1.868504,1.106345),\\
    (r_2^x,r_2^y)&=(-0.328201,0.562800),\\
    (r_3^x,r_3^y)&=(0.353471,0.112478),\\
    (r_4^x,r_4^y)&=(-0.301707,1.21919),
\end{split}
\end{align}
resulting in the following effective coupling matrix (rounded to 6dp)
\begin{equation}\label{eq:coupling_opt}
    J_{ij}=g^2\begin{bmatrix}
    0.900004& 0.099983& 0.000099& 0.000058\\0.099983& 
  1.066082& 0.099445& 0.000043\\ 0.000099& 0.099445& 
  1.096240& 0.099814\\ 0.000058& 0.000043& 0.099814& 
  0.900011
\end{bmatrix}.
\end{equation}
Note that this matrix has been scaled from the form given in Eq.~\eqref{eq:exp_coupling}, however using the unscaled form would only corresponds to a rescaling of the coupling strength $g$. In Fig.~\ref{fig:Jij_comparison} we show a comparison of the $J_{ij}^\mathrm{nn}$ elements with those of the above coupling matrix to highlight the difference in the coupling strengths, but also the similarity in the structure, with stronger couplings along the diagonal, and non-zero couplings in the first off-diagonal.

\bibliography{references_2} 

\end{document}